\documentclass[runningheads]{llncs}

\usepackage{accv}

\usepackage{accvabbrv}

\usepackage{graphicx}
\usepackage{booktabs}
\usepackage{placeins}
\usepackage{array}
\usepackage{multirow}
\usepackage{longtable}

\newif\ifincludeappendix
\includeappendixtrue

\makeatletter
\@ifundefined{pdfcompresslevel}{\newcount\pdfcompresslevel}{}
\@ifundefined{pdfoptionpdfminorversion}{\newcount\pdfoptionpdfminorversion}{}
\@ifundefined{pdfgentounicode}{\newcount\pdfgentounicode}{}
\@ifundefined{pdfglyphtounicode}{\newcommand{\pdfglyphtounicode}[2]{}}{}
\makeatother
\usepackage[accsupp]{axessibility}  % Improves PDF readability for those with disabilities.

\ifdefined\pdfoutput
\fi

\usepackage[hidelinks]{hyperref}
\hypersetup{
  pdftitle={TopoDiSE: Topology-Preserving Representations for Corneal IVCM Diagnosis},
  pdfauthor={Xueyao Wan and Hang Yu}
}

\usepackage{orcidlink}

\begin{document}

\title{TopoDiSE: Topology-Preserving Representations for Corneal IVCM Diagnosis} 

\titlerunning{TopoDiSE}

\author{Xueyao Wan\inst{1} \and Hang Yu\inst{2}}
\authorrunning{X. Wan and H. Yu}

\institute{Harbin Institute of Technology, China\\
\email{1185909349@qq.com}
\and
Nanyang Technological University, Singapore\\
\email{hang023@e.ntu.edu.sg}}

\maketitle
\begin{abstract}

In vivo corneal confocal microscopy (IVCM) provides a non-invasive view of
corneal small-fiber nerve morphology, but Parkinson's disease (PD) diagnostic pipelines 
often depend on handcrafted nerve statistics, segmentation-derived morphology, and 
image-level analysis in small cohorts. These designs standardize structure but can 
bottleneck patient-level representation learning by reducing continuous nerve topology 
to fixed masks or scalar summaries. We propose TopoDiSE, a
topol\-ogy-pre\-serv\-ing representation framework for data-efficient patient-level
analysis of Parkinson's disease from IVCM images.
TopoDiSE learns a topol\-ogy-pre\-serv\-ing semantic encoder from a broader IVCM pool
using nerve-mask-derived topology supervision, with pixel-space diffusion
reconstruction as a complementary constraint that keeps the representation tied
to full IVCM appearance, adapts it with
unsupervised region-spe\-cial\-ized continual pretraining for central/\allowbreak whorl organization, and
bridges image-level features to patient-level decisions by freezing the
original convolutional backbone, adding lightweight adapters, and aggregating
regional evidence with Atten\-tion-MIL and complementary morphology cues.
Across three repeats of patient-level five-fold validation, TopoDiSE
achieves accuracies of 0.850 $\pm$ 0.031, 0.598 $\pm$ 0.072, and
0.752 $\pm$ 0.057 for PD diagnosis, H\&Y-group prediction, and subtype screening,
respectively.
These results support topol\-ogy-pre\-serv\-ing IVCM representations
for small, regionally structured patient cohorts.
Code and experiment configurations are available at
\href{https://github.com/wanxueyao/topodise.git}{\nolinkurl{github.com/wanxueyao/topodise}}.


\end{abstract}

\section{Introduction}

In vivo corneal confocal microscopy (IVCM), also commonly referred to as corneal
confocal microscopy (CCM), offers a non-invasive, high-resolution view of the
corneal sub-basal nerve plexus. Because corneal small fibers reflect peripheral
and autonomic neurodegeneration, IVCM has been studied as an ophthalmic window
for neurological and metabolic diseases. In Parkinson's disease (PD), prior
studies have reported corneal nerve loss and altered nerve
microstructure~\cite{misra2017corneal,lim2020corneal}. Other work has linked
corneal nerve fiber loss to cognitive impairment and autonomic
involvement~\cite{che2021corneal,che2022corneal}, and to motor progression or
motor subtypes~\cite{lim2021corneal,che2023severity,li2025cornea}.
In diabetes mellitus (DM), IVCM has also been used to assess peripheral
neuropathy and to support automated diagnosis~\cite{preston2022artificial,chen2024development}.
These findings motivate IVCM-based diagnosis but leave an underexplored computer
vision question: how can transferable patient-level representations be learned
from small sets of fine nerve images instead of predefined morphometric
indicators?

Existing IVCM research leaves three gaps that are central to this paper. First,
many PD and DM studies reduce images to handcrafted or software-derived nerve
statistics such as density, length, branching, and fractal
dimension~\cite{misra2017corneal,lim2020corneal,che2020potential,chiang2023vivo},
or rely on segmentation-derived intermediates that can lose weak boundaries,
continuity uncertainty, and fine-scale intensity morphology. Second, automated IVCM methods have mostly targeted image
selection, segmentation, parameter estimation, or single-image classification
\cite{qiao2025automated,setu2022segmentation,essalat2023interpretable,zhang2026ivcm},
without directly modeling a patient's central and whorl-like images as a
diagnostic bag. Third, PD diagnosis, Hoehn and Yahr (H\&Y)-group prediction,
and Tremor Dominant (TD) versus non-TD subtype screening involve different
label structures and class distributions. This motivates a shared IVCM
representation with task-specific patient-level adaptation within the PD setting.

We propose \textbf{TopoDiSE}, a topology-preserving representation framework for
patient-level PD analysis from IVCM images. In Stage0, TopoDiSE learns a
diffusion semantic encoder from the CORN-1 IVCM image pool~\cite{imed2026corncomplex}
by combining pixel-space reconstruction with topology supervision from the
provided nerve-fiber segmentation masks. The mask is not the downstream
representation; it serves as a structural training signal, while diffusion keeps
the semantic vector tied to the full IVCM image. In Stage1, region-specialized
continual pretraining adapts the encoder to central/whorl organization before
labeled patient-level learning. In Stage2, the original convolutional backbone
is frozen. Residual adapters adapt the encoder, and attention-based multiple
instance learning (Attention-MIL) aggregates regional evidence for
task-specific patient-level inference.

Our contributions are threefold. First, we study patient-level topology
representations for PD diagnosis, H\&Y-group prediction, and TD/non-TD screening,
using mask-derived supervision to retain fine nerve structure beyond handcrafted
metrics or mask-only analysis. Second,
we introduce a staged pretraining protocol that connects external IVCM topology
learning with disease-label-free regional specialization before supervised
patient-level adaptation. Third, we develop a bounded adaptation and inference
protocol for small imbalanced cohorts with predefined candidate sets and
task-specific inference rules.

In repeated patient-level cross-validation, TopoDiSE achieved higher mean
accuracy and macro-F1 than the other image-based methods across all three PD
tasks. Its mean accuracy was close to that of the clinical-metric baseline
for PD diagnosis and TD/non-TD screening. In a separate transfer experiment,
we examine reuse of the pretrained representation on a
control-vs-diabetic retinopathy (DR) DM cohort.

\section{Related Work}

% Literature-selection rule for this paper: prioritize authoritative venues
% (Nature/Science family, npj, Scientific Reports, Communications Biology, and
% top CV/ML conferences such as CVPR/ICCV/ECCV/NeurIPS/ICML/ICLR), then recent
% work from the last three years and highly cited domain references. We do not
% cite weakly related general CV papers only to satisfy venue preference.

\paragraph{IVCM as a marker of neurodegeneration.}
IVCM provides a non-invasive window into the corneal sub-basal nerve plexus. A
growing body of ophthalmic and neurological evidence suggests that corneal nerve
morphology reflects small-fiber neurodegeneration in PD and related disorders.
Prior PD studies reported altered corneal nerve microstructure, automated
evidence of small-fiber loss~\cite{misra2017corneal,lim2020corneal}, links to
cognitive impairment, autonomic involvement, and differential diagnosis from
multiple system atrophy~\cite{che2021corneal,che2022corneal,niu2024corneal}, and
separation of heterogeneous motor subtypes, such as TD versus
non-TD~\cite{che2023severity,li2025cornea}. This evidence
establishes clinical relevance, but it is largely built on handcrafted indicators
such as density, length, width, branch density, and fractal dimension, rather
than a patient-level image-representation framework for diagnosis, H\&Y-group prediction, and
class-imbalanced subtype screening.

\paragraph{Automated IVCM image analysis.}
Automation is needed because IVCM acquisition, field selection, and manual
quantification are variable across studies~\cite{de2017laser,chiang2023vivo}.
Deep learning has broadly reshaped medical image analysis~\cite{litjens2017survey}
and has been used in IVCM for representative-image selection~\cite{qiao2025automated},
corneal nerve or dendritic-cell segmentation~\cite{setu2022segmentation}, diabetic
peripheral neuropathy (DPN) classification~\cite{preston2022artificial,chen2024development}, keratitis
classification~\cite{essalat2023interpretable}, and interactive IVCM
interpretation~\cite{zhang2026ivcm}. Recent CCM/IVCM studies have also moved
toward masked self-supervised transformers for corneal nerve segmentation and DPN
diagnosis, clinically guided CCM synthesis, and segmentation-map-free tortuosity
grading~\cite{zhang2025hmsvit,zhang2026wdlora,ouan2026selfsupervised}. These works demonstrate the value of
learning from IVCM images and the usefulness of explicit nerve structure, but
most diagnostic pipelines target image quality control, segmentation, or
single-image classification, and tend to treat regions as independent samples
despite acquisition protocols distinguishing locations within the corneal
sub-basal nerve plexus~\cite{de2017laser,chiang2023vivo}. This leaves open how to
use segmentation-derived structure as supervision while learning patient-level
representations from central and whorl-like observations.

\paragraph{Representation learning.}
General representation learning has advanced through contrastive, supervised
contrastive, masked, and large-scale self-supervised pretraining, including
Simple Framework for Contrastive Learning of Visual Representations (SimCLR),
supervised contrastive learning~\cite{chen2020simple,khosla2020supervised},
masked autoencoders (MAE)~\cite{he2022masked}, and DINOv2~\cite{oquab2023dinov2}.
Medical imaging has developed domain-aware pretraining such as Models
Genesis~\cite{zhou2019models} and segmentation foundation systems such as the
Segment Anything Model (SAM) and MedSAM~\cite{kirillov2023segment,ma2024segment}.
Diffusion models provide dense generative priors through denoising
reconstruction~\cite{ho2020denoising,nichol2021improved}, and recent medical
foundation-model studies emphasize data scaling~\cite{ilse2025datascaling} and
mask-guided self-supervision~\cite{gao2026learning}. Together, these
methods provide strong baselines, but they are not tailored to thin IVCM nerves:
when used alone, contrastive, masked, or diffusion objectives may lack the
topology-specific signal needed to decide which fine tubular structures should
be preserved, and segmentation foundation models focus on masks rather than
patient-level diagnostic representations. Attention-MIL supports bag
aggregation~\cite{ilse2018attention}, graph neural networks support relational
reasoning~\cite{hamilton2017inductive,velickovic2018graph}, adapter tuning
supports parameter-efficient transfer~\cite{chen2022adaptformer}, and
prototypical networks support few-shot classification~\cite{snell2017prototypical},
but they leave different gaps for IVCM: graph neural networks (GNNs) can reason over
topology but depend on explicit, error-prone node/edge construction, often from
lossy masks; MIL and adapter methods do not specify central/whorl organization.
The open problem is therefore to combine dense representation learning,
topology-aware supervision, and patient-level regional structure for small IVCM
cohorts.

\section{Methodology}

\begin{figure*}[t]
\centering
\includegraphics[width=0.98\textwidth]{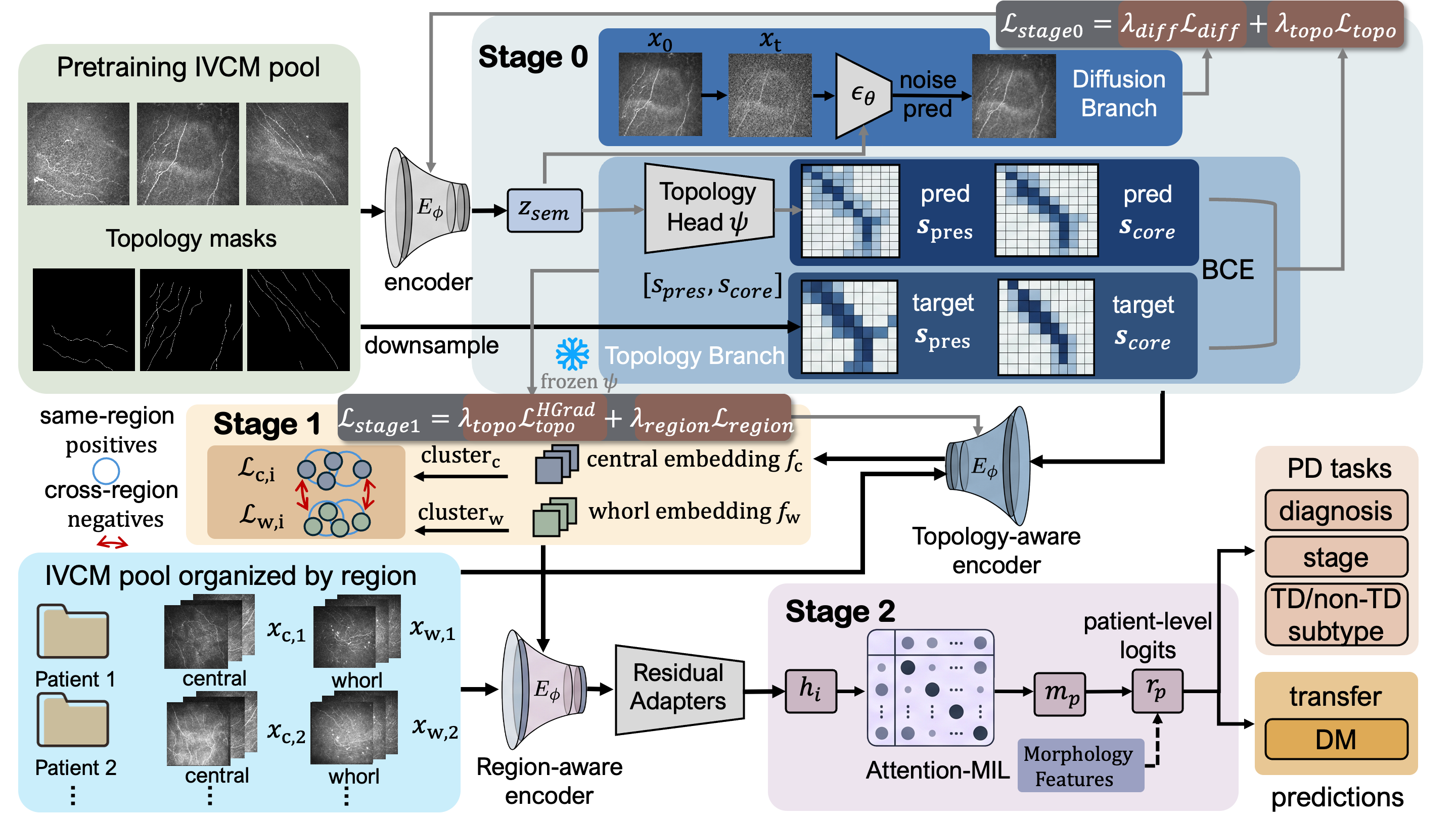}
\caption{\textbf{TopoDiSE overview.} Stage0 learns from CORN-1 images using pixel-space
noise prediction and nerve-mask topology targets. The diffusion panel
illustrates reconstruction from predicted noise. Stage1 adapts the encoder to
central and whorl regions using region contrast and dual-threshold HGrad targets
with the frozen topology head $\psi$. Stage2 starts from the Stage1 encoder and
uses residual adapters and Attention-MIL for patient-level prediction.}
\label{fig:topodise-overview}
\end{figure*}

As shown in Fig.~\ref{fig:topodise-overview}, TopoDiSE learns nerve structure
from masks in Stage0 and adapts the encoder to central and whorl regions in
Stage1. Stage2 uses patient-level diagnostic labels to adapt the model to each
downstream task.

\subsection{Problem Formulation}

Let $p$ index a patient. Each patient provides a set of central IVCM images
$X_p^c=\{x_{p,i}^c\}_{i=1}^{n_p^c}$ and whorl-like images
$X_p^w=\{x_{p,j}^w\}_{j=1}^{n_p^w}$. TopoDiSE represents the patient as a bag
of regional image pairs,
\begin{equation}
\mathcal{B}_p=\{(x_{p,i}^c,x_{p,j}^w):x_{p,i}^c\in X_p^c,
x_{p,j}^w\in X_p^w\},
\end{equation}
with random subsampling to a fixed maximum bag size when the Cartesian product is
large. The patient label satisfies $y_p\in\mathcal{Y}$, where the label space
$\mathcal{Y}$ depends on the downstream task.

The goal is to learn a representation function that maps the regional bag
$\mathcal{B}_p$ to a patient-level representation $r_p$, which is then used by
task-specific scoring and inference rules. Each PD task starts from the same Stage1 encoder and trains its own
adaptation and inference modules.

\subsection{Stage0: Topology-Preserving Diffusion Semantic Encoding}

The first stage learns a semantic vector $z_{\mathrm{sem}}$ from the external
CORN-1 IVCM image pool~\cite{imed2026corncomplex}, using nerve masks for structural
supervision without downstream disease labels. Unlike generic image pretraining objectives,
this stage is explicitly biased toward thin nerve topology. Because CORN-1 does
not provide patient-level central/whorl pairs, Stage0 forms two independently augmented views
$x^{(1)}$ and $x^{(2)}$ from the same IVCM image and feeds them as
pseudo-branch inputs to the dual-branch encoder $E_\phi$, which produces
\begin{equation}
z_{\mathrm{sem}}=E_\phi(x^{(1)},x^{(2)})\in\mathbf{R}^d,
\end{equation}
where $d=512$ in our implementation. The encoder uses two region branches with
GroupNorm-based residual blocks and a cross-view fusion module, avoiding
BatchNorm statistics that are unstable in small medical batches.

TopoDiSE supervises $z_{\mathrm{sem}}$ with two complementary signals. The first
is pixel-space conditional diffusion reconstruction, following denoising
diffusion probabilistic models~\cite{ho2020denoising,nichol2021improved}. We
independently sample a clean reference augmentation $x_0$ from the same source
image. For a timestep $t$ and Gaussian noise $\epsilon$, the noised image is
\begin{equation}
x_t=\sqrt{\bar{\alpha}_t}x_0+\sqrt{1-\bar{\alpha}_t}\epsilon.
\end{equation}
A conditional denoising network $\epsilon_\theta$ predicts the injected noise
from $(x_t,t,z_{\mathrm{sem}})$, yielding
\begin{equation}
\mathcal{L}_{\mathrm{diff}}=
\|\epsilon-\epsilon_\theta(x_t,t,z_{\mathrm{sem}})\|_2^2.
\end{equation}
Here $z_{\mathrm{sem}}$ conditions the denoising blocks through adaptive group
normalization rather than being noised itself, so reconstruction requires
semantic information for pixel-space nerve structure. We reconstruct in pixel
space rather than a variational autoencoder (VAE) latent because coarse
compression can erase fine-scale IVCM nerve fibers.

The topology head $\psi$ projects $z_{\mathrm{sem}}$ onto a $4\times4\times128$
feature grid and upsamples it with lightweight convolutional blocks to predict
nerve presence and dense core on a $G\times G$ grid ($G=32$). In Stage0, we
construct the target $s=[s_{\mathrm{pres}},s_{\mathrm{core}}]$ from the CORN-1
nerve mask $M$. We apply the same geometric augmentation to $M$ and $x_0$,
using nearest-neighbor interpolation for the mask. For each grid cell
$\Omega_{ij}$, we compute the nerve density as
\begin{equation}
d_{ij}=\frac{1}{|\Omega_{ij}|}\sum_{(u,v)\in\Omega_{ij}} M_{uv}.
\end{equation}
The two target channels are then defined as
\begin{equation}
s_{\mathrm{pres},ij}=\mathbb{I}[d_{ij}>0],\qquad
s_{\mathrm{core},ij}=\mathbb{I}[d_{ij}>\mu_d+\sigma_d],
\end{equation}
where $\mu_d$ and $\sigma_d$ are the mean and standard deviation of
$\{d_{ij}\}$ within the same image. Thus $s_{\mathrm{pres}}$ indicates whether a
local patch contains any nerve fiber, while $s_{\mathrm{core}}$ highlights patches with
high pooled mask density.
With topology target $s$, the topology loss uses binary cross-entropy (BCE),
\begin{equation}
\mathcal{L}_{\mathrm{topo}}=\mathrm{BCE}(\psi(z_{\mathrm{sem}}),s).
\end{equation}
The mask-derived objective encourages recoverable local nerve-structure
information in the semantic vector, while pixel-space noise prediction provides
a complementary full-image appearance objective. The complete Stage0 objective is
\begin{equation}
\mathcal{L}_{\mathrm{stage0}}=
\lambda_{\mathrm{diff}}^{(0)}\mathcal{L}_{\mathrm{diff}}+
\lambda_{\mathrm{topo}}^{(0)}\mathcal{L}_{\mathrm{topo}},
\end{equation}
where the superscript denotes the training stage.

\subsection{Stage1: Region-Specialized Continual Pretraining}

Stage1 adopts the central/whorl organization of established clinical IVCM
acquisition protocols. Since CORN-1 pretraining uses augmented views of individual
images rather than patient-level central/whorl pairs, region-specialized continual
pretraining (RSCP) adapts the encoder to this regional input structure without
disease labels. It continues from the Stage0 encoder, updating the later branch
stages, cross-region fusion and projection layers, and GroupNorm affine parameters
in the earlier stages. The topology-head parameters remain frozen, while gradients
through the head update the trainable encoder parameters. The diffusion loss is
excluded from the default Stage1 objective.

Stage1 uses a disease-label-free IVCM pool organized by subject and region, kept
outside all Stage2 supervised fold construction and evaluation. Subject grouping
is used only to assemble valid regional inputs, while contrastive supervision uses
branch region identity rather than diagnostic or patient labels. Paired subjects
provide valid central/whorl inputs; single-region subjects use duplicated inputs
only for topology regularization and are masked out of the region loss. A mini-batch
contains tensors $X_c$ and $X_w$, which produce central embeddings
$\{f_{c,i}\}$, whorl-like embeddings $\{f_{w,i}\}$, and fused semantic vectors
$z_i$.
Because the Stage1 regional pool lacks nerve masks, we derive dual-threshold
horizontal-gradient (HGrad) targets from the augmented central input. Absolute
horizontal pixel differences are zero-padded on the left and average-pooled to a
$32\times32$ response map $R$. The two target channels are
\begin{equation}
s_{\mathrm{H},0}=\mathbb{I}[R>\mu_R],\qquad
s_{\mathrm{H},1}=\mathbb{I}[R>\mu_R+\sigma_R],
\end{equation}
where $\mu_R$ and $\sigma_R$ are the mean and sample standard deviation within
each pooled response map. The Stage1 topology loss compares $\psi(z)$ with these
image-derived targets using BCE.

Central and whorl-like regions remain anatomically different, so the region
objective is a region-level regularizer rather than a patient-identity objective.
It is computed only on valid paired samples, where
same-region terms form the numerator and cross-region terms form the competing
denominator. For a central anchor,
\begin{equation}
\mathcal{L}_{c,i}=-\log
\frac{\sum_{j\ne i}\exp(\mathrm{sim}(f_{c,i},f_{c,j})/\tau)}
{\sum_{j\ne i}\exp(\mathrm{sim}(f_{c,i},f_{c,j})/\tau)+
 \sum_k \exp(\mathrm{sim}(f_{c,i},f_{w,k})/\tau)}.
\end{equation}
The whorl-like side $\mathcal{L}_{w,i}$ is defined symmetrically, and
$\mathcal{L}_{\mathrm{region}}$ averages the two directions over the batch.
The Stage1 objective is
\begin{equation}
\mathcal{L}_{\mathrm{stage1}}=
\lambda_{\mathrm{region}}^{(1)}\mathcal{L}_{\mathrm{region}}+
\lambda_{\mathrm{topo}}^{(1)}\mathcal{L}_{\mathrm{topo}}.
\end{equation}
Stage1 uses a smaller topology weight than Stage0, using HGrad as an auxiliary
structural constraint while adapting the encoder to regional organization before
patient-level supervised training.

\subsection{Stage2: Bounded Patient-Level Adaptation}

\begin{figure*}[t]
\centering
\includegraphics[width=0.98\textwidth]{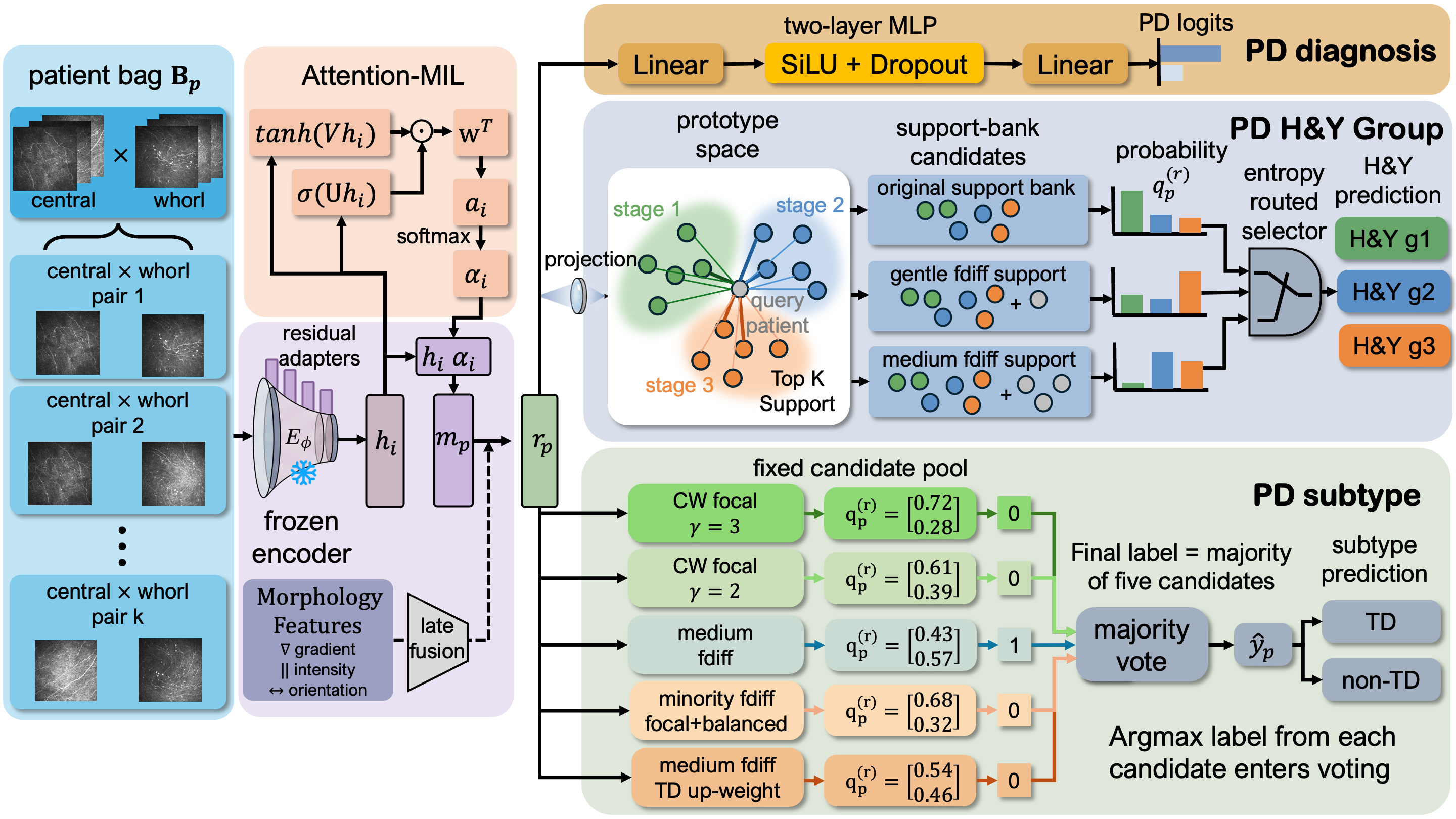}
\caption{\textbf{Stage2 patient-level adaptation and task-specific inference.}
Residual adapters are trained with the original convolutional backbone frozen.
Gated Attention-MIL pools regional-pair embeddings into $m_p$. PD diagnosis
and H\&Y-group prediction fuse $m_p$ with morphology features, while subtype
screening uses $m_p$ directly. PD diagnosis uses a two-layer MLP, H\&Y-group
prediction selects the support log-sum-exp (support-LSE) candidate with minimum
predictive entropy, and subtype screening takes the majority vote of five
candidate labels.}
\label{fig:stage2-adaptation}
\end{figure*}

Stage2 freezes the original convolutional backbone parameters and trains residual
adapters, Attention-MIL and task-specific heads. Cross-region fusion and projection
are fine-tuned for PD diagnosis, H\&Y-group prediction and DM transfer, and remain
fixed for subtype screening. For a feature map $u$, an adapter has the residual form
\begin{equation}
A(u)=u+W_{\mathrm{up}}\,\rho(W_{\mathrm{down}}u),
\end{equation}
where $\rho$ is the Gaussian error linear unit (GELU), $W_{\mathrm{down}}$ reduces channel dimension, and
$W_{\mathrm{up}}$ is near-zero initialized~\cite{chen2022adaptformer}.

Each regional pair in the patient bag $\mathcal{B}_p$ is encoded by the
Stage1-initialized encoder with adapters, yielding a fused instance embedding $h_i$, and
the bag of instance embeddings is aggregated with gated
Attention-MIL~\cite{ilse2018attention}:
\begin{equation}
a_i=w^\top\big(\tanh(Vh_i)\odot\sigma(Uh_i)\big),
\quad
\alpha_i=\frac{\exp(a_i)}{\sum_j\exp(a_j)},
\quad
m_p=\sum_i\alpha_i h_i.
\end{equation}
Here $w$, $V$, and $U$ are learnable attention parameters, and $m_p$ is the
patient embedding obtained without assigning labels to individual images.

TopoDiSE can fuse a morphology side channel complementary to $m_p$. Let
$\mathrm{stat}(\cdot)$ denote intensity, gradient, high-gradient fraction, and
anisotropy statistics; for each regional pair, these statistics and their
absolute difference are averaged over the bag:
\begin{equation}
g_p=\frac{1}{|\mathcal{B}_p|}\sum_{(x^c,x^w)\in\mathcal{B}_p}
[\,\mathrm{stat}(x^c)\Vert \mathrm{stat}(x^w)\Vert
|\mathrm{stat}(x^c)-\mathrm{stat}(x^w)|\,].
\end{equation}
A lightweight projection maps the 21-dimensional vector $g_p$ to a morphology
embedding, which is concatenated with $m_p$ to form
\begin{equation}
r_p=[\,m_p\Vert \eta(g_p)\,].
\end{equation}
Morphology fusion is used for PD diagnosis, H\&Y-group prediction and DM transfer;
subtype screening uses $r_p=m_p$.

Stage2 combines patient classification with task-dependent auxiliary losses:
\begin{equation}
\mathcal{L}_{\mathrm{stage2}}=
\mathcal{L}_{\mathrm{cls}}+
\lambda_{\mathrm{proto}}\mathcal{L}_{\mathrm{proto}}+
\lambda_{\mathrm{fdiff}}\mathcal{L}_{\mathrm{fdiff}}+
\lambda_{\mathrm{aug}}\mathcal{L}_{\mathrm{aug}}.
\end{equation}
Here, $\mathcal{L}_{\mathrm{cls}}$ is the patient classification loss
(cross-entropy or focal loss), $\mathcal{L}_{\mathrm{proto}}$ supervises
projected H\&Y patient embeddings, $\mathcal{L}_{\mathrm{fdiff}}$ trains the
feature diffusion head, and $\mathcal{L}_{\mathrm{aug}}$ applies the same
classification loss to augmented patient bags.

For a clean pair embedding $h_0$, the diffusion head reconstructs $h_0$ from
$h_t=\sqrt{\bar{\alpha}_t}h_0+\sqrt{1-\bar{\alpha}_t}\xi$ by minimizing
$\mathcal{L}_{\mathrm{fdiff}}=\|D_\omega(h_t,t,y)-h_0\|_2^2$.
Reconstructed pair embeddings are softly blended with their source embeddings
according to cosine similarity, then pooled by Attention-MIL to form augmented
patient bags. Candidate-specific use of feature diffusion and support
augmentation is summarized in Appendix~C.

\subsection{Task-Specific Patient-Level Inference}

The H\&Y and subtype candidates are trained independently from the same Stage1
initialization, and their predictions are combined at inference.

For binary diagnosis, a two-layer MLP classifier maps $r_p$ to patient-level
logits. For H\&Y-group prediction, TopoDiSE uses a support-bank prototype head related
to prototypical classification~\cite{snell2017prototypical}. We project and normalize
the patient representation,
\begin{equation}
e_p=\frac{W_{\mathrm{proto}}r_p}{\|W_{\mathrm{proto}}r_p\|_2},
\end{equation}
and compare it with training-fold support embeddings for each H\&Y group. For
class $c$, the support set $S_c$ contains training-patient embeddings and their
augmented counterparts where enabled. The support-LSE logit is
\begin{equation}
\ell_{p,c}^{\mathrm{proto}}=
\log\left(\frac{1}{K_c}\sum_{j\in\mathrm{TopK}_{K_c}(S_c;e_p)}
\exp\left(-\frac{\|e_p-e_{c,j}\|_2^2}{\tau}\right)\right).
\end{equation}
Here $\mathrm{TopK}_{K_c}(S_c;e_p)$ selects the nearest support embeddings in class
$c$ under squared Euclidean distance, with \mbox{$K_c=\min(5,|S_c|)$}.
The final H\&Y prediction uses a fixed support-LSE candidate set with entropy routing.
For patient $p$, let $q_p^{(r)}$ denote the probability vector from
candidate $r$. TopoDiSE selects the candidate and label as
\begin{equation}
r^*(p)=\arg\min_r H\big(q_p^{(r)}\big),\qquad
\hat y_p=\arg\max_c q_{p,c}^{(r^*(p))},
\end{equation}
where $H(q)=-\sum_c q_c\log q_c$.

For subtype screening, TopoDiSE uses five candidates with different
class-weighting, focal-loss~\cite{lin2017focal}, and feature-diffusion
configurations (Appendix~C). Each candidate contributes one TD/non-TD argmax
label, and the final prediction is the majority vote.

\section{Experiments}

\subsection{Experimental Setup}

\subsubsection{Datasets and Tasks}
The PD cohort of Li et al.~\cite{li2025cornea} contains 434 central and
whorl-like IVCM images from 63 PD patients and 51 controls (31 cataract and
20 healthy). Diagnosis (\texttt{pd\_binary}) uses 110 patients with paired
regions, as four healthy controls have only whorl images. Among PD patients,
\texttt{pd\_stage} distinguishes three H\&Y groups (22/30/11 patients) and
\texttt{pd\_subtype} distinguishes TD from non-TD (23/40).
Stage0 uses external CORN-1 images and their annotated nerve masks~\cite{imed2026corncomplex}.
After excluding filename matches with the downstream PD and DM datasets,
Stage1 used 219 central--whorl pairs and 344 central-only samples. All samples
contributed to the structural loss, and only genuine pairs contributed to the
regional contrastive loss. The separate DM transfer cohort contains 614 images from 20 controls and
60 DR subjects (\texttt{dm\_binary}).

\subsubsection{Metrics and Model Selection}
We used patient-level five-fold validation repeated three times. Table~\ref{tab:pd-main-results}
reports mean accuracy (primary) and macro-F1 with 95\% interval half-widths
over the 15 fold scores. Repeats share patients, so these intervals summarize
dispersion. DM transfer also reports balanced accuracy (BAcc.) and area under
the ROC curve (AUC). For PD, each validation fold serves for checkpoint
selection by a predefined metric and for evaluation. Candidates and decision
rules were fixed in advance: one classifier for diagnosis, entropy routing for
H\&Y and majority voting for subtype.

\subsubsection{Baselines}
We compare with the following image-based baselines: an ImageNet-initialized dual-stream
\textbf{ResNet18}~\cite{he2016deep}; \textbf{SimCLR} contrastive pretraining
with a ResNet18 dual-stream backbone followed by a frozen-backbone downstream
classifier~\cite{chen2020simple}; \textbf{UNet-Only}, which first applies a
CORN-1-trained U-Net segmentation model~\cite{ronneberger2015u} to produce nerve
maps and then classifies only these segmentation-derived maps with a CNN;
\textbf{UNet-Dual}, which uses the same U-Net maps but adds a raw-IVCM branch for
dual-view CNN fusion; a skeleton \textbf{GNN}, which skeletonizes the U-Net nerve
maps, converts endpoints, branch points, and fiber segments into a graph, and
encodes it with a graph attention network (GAT)~\cite{velickovic2018graph};
\textbf{MTL/MAE}, a multi-task learning
(MTL) baseline whose auxiliary reconstruction branch is masked autoencoding
(MAE)~\cite{he2022masked}; and episodic
\textbf{Prototype} learning~\cite{snell2017prototypical}, which uses a dual-stream
ResNet18 encoder and class prototypes formed from labeled support embeddings.
After episodic training, validation patients are classified against global
training-fold prototypes.

\textbf{Nerve-LR} uses 11 corneal nerve metrics: CNFD, CNBD, CNFL, CTBD, CNFA, CNFW, CNFractal, IWL, IWFA, IWFW,
and IWFrD. Measurements are averaged per patient. Class-balanced L2 logistic regression uses
training-fold median imputation and standardization. Three-fold stratified inner
validation selects $C\in\{0.01,0.1,1,10\}$ by macro-F1, with preprocessing
refitted in each inner training fold.

\subsubsection{External Data and Fairness Protocol}
Baselines use their strongest predefined valid configurations. SimCLR reports
the stronger of PD training-fold and CORN-1 self-supervised pretraining for
each task. The three nerve-map baselines use CORN-1 annotations through the
U-Net, and MTL/MAE uses CORN-1 images for reconstruction. Episodic Prototype
uses labeled training-fold supports. TopoDiSE uses CORN-1 masks directly,
followed by the disease-label-free regional stage. Section~\ref{sec:mechanism-analysis}
tests the contribution of each stage.

\subsection{Main PD Results}

\begin{table*}[t]
\centering
\caption{\textbf{Patient-level PD comparison.} Accuracy and macro-F1 are fold
means $\pm$ 95\% CI half-width over 5-fold $\times$ 3 validation. Diagnosis
subsets are marked $^\dagger$ (94 patients) and $^\ddagger$ (110 patients).}
\label{tab:pd-main-results}
\scriptsize
\setlength{\tabcolsep}{2.7pt}
\resizebox{\textwidth}{!}{%
\begin{tabular}{lcccccc}
\toprule
\textbf{Method} & \multicolumn{2}{c}{\textbf{\texttt{pd\_binary}}} & \multicolumn{2}{c}{\textbf{\texttt{pd\_stage}}} & \multicolumn{2}{c}{\textbf{\texttt{pd\_subtype}}} \\
\cmidrule(lr){2-3}\cmidrule(lr){4-5}\cmidrule(lr){6-7}
& \textbf{Acc.} & \textbf{F1} & \textbf{Acc.} & \textbf{F1} & \textbf{Acc.} & \textbf{F1} \\
\midrule
ResNet18 & $0.805\pm0.036$ & $0.792\pm0.042$ & $0.457\pm0.070$ & $0.229\pm0.050$ & $0.651\pm0.029$ & $0.475\pm0.070$ \\
SimCLR & $0.775\pm0.031$ & $0.754\pm0.040$ & $0.510\pm0.045$ & $0.277\pm0.045$ & $0.697\pm0.029$ & $0.540\pm0.082$ \\
UNet-Only & $0.696\pm0.032$ & $0.673\pm0.036$ & $0.499\pm0.045$ & $0.247\pm0.039$ & $0.641\pm0.015$ & $0.401\pm0.024$ \\
UNet-Dual & $0.752\pm0.031$ & $0.746\pm0.033$ & $0.442\pm0.044$ & $0.240\pm0.041$ & $0.646\pm0.022$ & $0.410\pm0.042$ \\
GNN & $0.799\pm0.017$ & $0.444\pm0.005$ & $0.475\pm0.041$ & $0.225\pm0.029$ & $0.636\pm0.013$ & $0.389\pm0.005$ \\
MTL/MAE & $0.820\pm0.045$ & $0.807\pm0.054$ & $0.399\pm0.079$ & $0.238\pm0.058$ & $0.668\pm0.039$ & $0.491\pm0.083$ \\
Prototype & $0.736\pm0.036$ & $0.726\pm0.037$ & $0.431\pm0.055$ & $0.328\pm0.042$ & $0.540\pm0.059$ & $0.494\pm0.072$ \\
Nerve-LR & $0.859\pm0.038^{\dagger}$ & $0.846\pm0.040^{\dagger}$ & $0.407\pm0.065$ & $0.392\pm0.068$ & $\mathbf{0.767\pm0.052}$ & $\mathbf{0.743\pm0.056}$ \\
\textbf{TopoDiSE} & $0.850\pm0.031^{\ddagger}$ & $0.844\pm0.035^{\ddagger}$ & $\mathbf{0.598\pm0.072}$ & $\mathbf{0.569\pm0.084}$ & $0.752\pm0.057$ & $0.669\pm0.103$ \\
\bottomrule
\end{tabular}
}
\end{table*}

TopoDiSE achieved higher mean accuracy and macro-F1 than all other image-based
methods in Table~\ref{tab:pd-main-results} across the three PD tasks. Its mean
accuracy was close to that of Nerve-LR for both PD diagnosis and TD/non-TD
screening.

For TopoDiSE, the companion BAcc/AUC diagnostics are $0.845/0.901$ on binary,
$0.569/0.658$ on H\&Y, and $0.709/0.659$ on subtype. The fine-grained tasks
have wider accuracy and macro-F1 intervals.

\subsection{Transferability to DM}

\begin{table}[t]
\centering
\caption{\textbf{DM transfer validation.} Values are patient-level means over 15 folds. Accuracy is shown with its reported 95\% interval half-width, and the other metrics with their fold standard deviations.}
\label{tab:dm-transfer}
\scriptsize
\setlength{\tabcolsep}{3pt}
\begin{tabular}{lcccc}
\toprule
\textbf{Method} & \textbf{Acc.} & \textbf{BAcc.} & \textbf{Macro-F1} & \textbf{AUC} \\
\midrule
ResNet18 & $0.783\pm0.029$ & $0.589\pm0.093$ & $0.573\pm0.139$ & $0.807\pm0.060$ \\
SimCLR & $0.783\pm0.032$ & $0.567\pm0.114$ & $0.528\pm0.159$ & $0.694\pm0.251$ \\
MTL/MAE & $\mathbf{0.813\pm0.039}$ & $0.708\pm0.093$ & $0.718\pm0.098$ & $\mathbf{0.833\pm0.083}$ \\
\textbf{TopoDiSE} & $\mathbf{0.813\pm0.046}$ & $\mathbf{0.847\pm0.068}$ & $\mathbf{0.787\pm0.084}$ & $0.818\pm0.104$ \\
\bottomrule
\end{tabular}
\end{table}

Because the DM images lack region labels, we feed the same image to both encoder branches.
The patient-level adaptation modules are trained on the DM folds. Table~\ref{tab:dm-transfer}
shows that TopoDiSE matches the strongest DM accuracy baseline while improving
BAcc and macro-F1 over MTL/MAE, indicating a less DR-majority-dominated
operating point. Its AUC is comparable to those of the other transfer models.

\subsection{Ablation Studies}
\label{sec:mechanism-analysis}

\begin{table}[!htb]
\centering\small
\caption{\textbf{Supervision targets across pretraining stages.} Accuracy and macro-F1 use the same 15 folds and final decision rules as Table~\ref{tab:pd-main-results}.}
\label{tab:mechanism-ablation}
\setlength{\tabcolsep}{3pt}
\begin{tabular*}{\linewidth}{@{\extracolsep{\fill}}llrrrrrr@{}}
\toprule
\multicolumn{2}{c}{Target} & \multicolumn{2}{c}{Diagnosis} & \multicolumn{2}{c}{H\&Y grouping} & \multicolumn{2}{c}{TD/non-TD} \\
\cmidrule(r){1-2}\cmidrule(lr){3-4}\cmidrule(lr){5-6}\cmidrule(l){7-8}
Stage0 & Stage1 & Acc. & F1 & Acc. & F1 & Acc. & F1 \\
\midrule
GT     & HGrad  & 0.850 & 0.844 & 0.598 & 0.569 & 0.752 & 0.669 \\
\addlinespace[3pt]
Frangi & HGrad  & 0.773 & 0.748 & 0.513 & 0.464 & 0.721 & 0.631 \\
GT     & Frangi & 0.851 & 0.844 & 0.585 & 0.521 & 0.636 & 0.399 \\
Frangi & Frangi & 0.777 & 0.756 & 0.536 & 0.489 & 0.671 & 0.537 \\
\bottomrule
\end{tabular*}
\end{table}

\paragraph{Pretraining targets.}
TopoDiSE uses annotated nerve masks (GT) in Stage0 and dual-threshold HGrad
in Stage1 (Table~\ref{tab:mechanism-ablation}). Replacing only Stage0 GT with
Frangi reduced macro-F1 on all three PD tasks. Replacing only Stage1 HGrad
with Frangi reduced macro-F1 from 0.569 to 0.521 for H\&Y and from 0.669 to
0.399 for subtype, while diagnosis scored 0.844 in both settings. The supplement
reports all six combinations.

\begin{figure}[!htb]
\centering
\includegraphics[width=\linewidth]{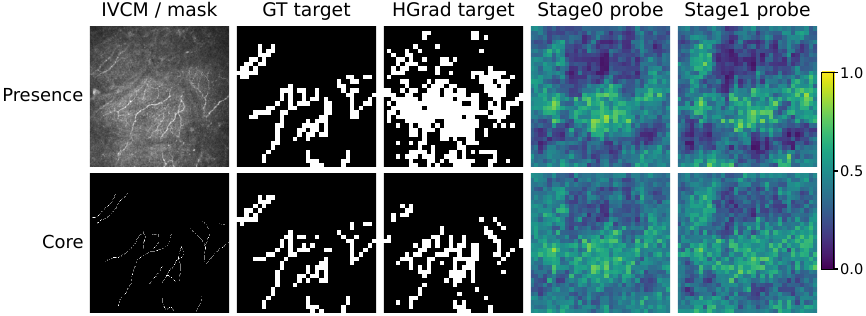}
\caption{\textbf{Targets and frozen-probe predictions on one CORN-1 image.}
Rows show GT presence/core, low/high HGrad thresholds and the corresponding presence/core probabilities from frozen Stage0 and Stage1 probes (seed 42). The first column shows the image and its annotation.}
\label{fig:targets-probes}
\end{figure}

\paragraph{What structure does the encoder retain?}
At the same diffusion training budget, Stage0 mask supervision increased
mean per-image average precision (AP) on 125 CORN-1 images by 0.108--0.124
for nerve presence and 0.066--0.078 for core across three seeds.
Continuing to Stage1 changed AP by $-0.0037$ and $-0.0027$ at seed 42
(Appendix~F.2). In the first prespecified example (Fig.~\ref{fig:targets-probes}),
HGrad responds to some edges outside the mask, while both probes recover
a similar coarse nerve pattern.

\begin{table}[!htb]
\centering\small
\caption{\textbf{Component ablations for one fixed candidate per task.}
Macro-F1 is computed on pooled out-of-fold predictions per repeat, then averaged
over three repeats and three pretraining seeds. Dashes mark inapplicable comparisons.}
\label{tab:component-ablation}
\setlength{\tabcolsep}{3pt}
\begin{tabular*}{\linewidth}{@{\extracolsep{\fill}}lrrr@{}}
\toprule
 & Diagnosis & H\&Y & Subtype \\
Configuration & US0 & U1 & TDREST5 \\
\midrule
Full candidate (GT+HGrad) & 0.846 & 0.524 & 0.625 \\
\midrule
\multicolumn{4}{@{}l}{\itshape Pretraining} \\
Without Stage0 image diffusion & 0.793 & 0.548 & 0.671 \\
Without Stage1 HGrad loss & 0.842 & 0.562 & 0.650 \\
Without both topology losses & 0.742 & 0.500 & 0.644 \\
Without Stage1 regional loss & 0.834 & 0.557 & 0.645 \\
Skip Stage1 & 0.832 & 0.543 & 0.722 \\
\midrule
\multicolumn{4}{@{}l}{\itshape Patient-level adaptation} \\
Without residual adapters & 0.831 & 0.487 & 0.630 \\
Fine-tune full encoder (fixed schedule) & 0.721 & 0.531 & 0.512 \\
Mean pooling in place of attention & 0.843 & 0.520 & 0.630 \\
Without morphology fusion & 0.837 & 0.518 & -- \\
Image morphology alone & 0.408 & 0.410 & 0.677 \\
Without prototype module & -- & 0.453 & -- \\
Without feature diffusion & -- & 0.476 & 0.660 \\
\midrule
\multicolumn{4}{@{}l}{\itshape Inference with fixed U1 checkpoints} \\
Classifier only & -- & 0.373 & -- \\
One mean prototype per class & -- & 0.410 & -- \\
Without generated supports & -- & 0.460 & -- \\
\bottomrule
\end{tabular*}
\end{table}

\paragraph{Component ablations.}
Table~\ref{tab:component-ablation} compares each variant with its full GT+HGrad
candidate: US0 for diagnosis, U1 for H\&Y and TDREST5 for subtype. Removing
Stage0 diffusion lowered US0 macro-F1 from 0.846 to 0.793 (paired
interval in Appendix~E.2). Without Stage1 HGrad, removing Stage0 mask
supervision lowered US0 from 0.842 to 0.742. Removing the regional loss
alone lowered US0 to 0.834. Skipping Stage1 raised TDREST5 from 0.625 to 0.722.
In a separate matched-view comparison, paired inputs scored highest for US0,
and central-only inputs scored highest for U1 and TDREST5 (Appendix~E.3).

\paragraph{Patient adaptation and morphology.}
Full encoder fine-tuning at the same schedule scored below adapters on
diagnosis and subtype, and slightly higher on H\&Y. Removing adapters reduced
U1 macro-F1 to 0.487, while mean pooling gave 0.520. Morphology fusion
added little to US0 or U1. Image-derived morphology alone scored lower on both
tasks but exceeded TDREST5 on subtype (0.677 versus 0.625).

\paragraph{Support-based inference and subtype imbalance.}
Training U1 without the prototype module or feature diffusion reduced macro-F1
from 0.524 to 0.453 or 0.476, respectively (Table~\ref{tab:component-ablation}).
With the full U1 checkpoints fixed, classifier-only inference scored 0.373
and removing generated supports gave 0.460. For TDREST5, disabling diffusion
raised macro-F1 from 0.625 to 0.660 but lowered TD recall from 0.729 to 0.638 (Appendix~E.5).
Appendix~E.2 and E.5 detail the class-weight and focal-loss comparisons.

\begin{table}[!htb]
\centering\small
\caption{\textbf{Patient-level decision rules.} GT+HGrad predictions and 15-fold means follow Table~\ref{tab:pd-main-results}. The best candidate has the highest mean macro-F1 for its task.}
\label{tab:decision-ablation}
\setlength{\tabcolsep}{4pt}
\begin{tabular*}{\linewidth}{@{\extracolsep{\fill}}lrrrr@{}}
\toprule
 & \multicolumn{2}{c}{H\&Y grouping} & \multicolumn{2}{c}{TD/non-TD} \\
\cmidrule(lr){2-3}\cmidrule(l){4-5}
Prediction rule & Acc. & Macro-F1 & Acc. & Macro-F1 \\
\midrule
Best individual candidate & 0.569 & 0.526 & 0.720 & 0.608 \\
Mean probability & 0.593 & 0.546 & 0.625 & 0.571 \\
Task-specific decision & 0.598 & 0.569 & 0.752 & 0.669 \\
\bottomrule
\end{tabular*}
\end{table}

\paragraph{Patient-level decisions.}
With GT+HGrad candidate predictions fixed (Table~\ref{tab:decision-ablation}),
replacing minimum-entropy routing with probability averaging reduced H\&Y
macro-F1 from 0.569 to 0.546. Replacing majority voting with probability
averaging reduced subtype macro-F1 from 0.669 to 0.571.

\section{Discussion and Conclusion}

Across the three PD tasks, TopoDiSE achieved higher mean accuracy and macro-F1
than all other image-based methods in our comparison. Its mean accuracy was
close to that of Nerve-LR for both PD diagnosis and TD/non-TD screening. The frozen probes show that Stage0 mask supervision makes nerve presence and
dense-core structure easier to recover. With this supervision fixed, Stage1
HGrad improves mean macro-F1 over Frangi on both fine-grained PD tasks.
Stage2 feature diffusion helps the H\&Y support-based candidate, while its
effect on subtype depends on the balance between TD and non-TD recall.

The separate DM transfer experiment supports reuse of an IVCM encoder across
clinical tasks. The staged design allows sites to train patient-level modules
locally after pretraining on external masks and unlabeled regional images.
Larger multi-center cohorts with harmonized acquisition and prospective
evaluation are needed to establish how well these findings generalize.
Reducing diffusion cost and reliance on annotated masks remains a practical goal.

\subsubsection*{Acknowledgements.}
We thank Dr. Xin Jin (Eye Hospital, The First Affiliated Hospital of Harbin
Medical University) for collaboration on the IVCM dataset, and Prof. Jianxing Liu
and Prof. Ligang Wu (Harbin Institute of Technology) for their guidance.

\clearpage
\ifincludeappendix
\appendix
\renewcommand{\theHsection}{supp.\Alph{section}}
\renewcommand{\theHtable}{supp.\arabic{table}}
\renewcommand{\theHfigure}{supp.\arabic{figure}}
\raggedbottom
\setcounter{table}{0}
\renewcommand{\thetable}{S\arabic{table}}
\setcounter{figure}{0}
\renewcommand{\thefigure}{S\arabic{figure}}
\begin{center}
{\Large\bfseries Supplementary Material for TopoDiSE\par}
\end{center}
\vspace{1em}
\suppressfloats[t]
\section{Data sources and patient splits}
\label{app:data-audit}

\subsection{Sources and regional pairing}
Stage0 used 1,516 CORN-1 images and their nerve masks. The Stage1 source
manifest contained 52 subjects, including 13 with both regions and 39 with
central images only. The 582 input rows comprised 234 central--whorl pairs
and 348 central-only samples. Filename matches with the PD and DM datasets
excluded 15 paired and four central-only rows, leaving 563 usable rows
(219 paired and 344 central-only). Only
the paired rows contributed to the regional contrastive loss.

The PD cohort contains 63 PD patients, 31 cataract controls and 20 healthy
controls~\cite{li2025cornea}. Four healthy controls have whorl images only,
leaving 110 patients for paired-image diagnosis. Nerve-LR uses 94 patients
with eligible clinical measurements for diagnosis. H\&Y grouping (22/30/11)
and TD/non-TD screening (23/40) use the 63 PD patients. The cohort labels
combine PIGD and mixed phenotypes in the non-TD class.

\subsection{Subject and image overlap}
The 563 Stage1 rows used 782 distinct image paths. We compared these with the
full PD and DM manifests (884 and 614 image paths, respectively).
Table~\ref{tab:app-image-overlap} summarizes exact and near-duplicate matches.
The Stage1 and PD manifests share seven subject names. The PD and DM manifests
share 20 subject names and 144 identical image pairs, so we report DM as a
transfer study. Acquisition session, medication and ocular comorbidity were
unavailable for adjustment.

\begin{table}[!htbp]
\centering\small
\caption{Image overlap across manifests. Exact matches have identical SHA-256
hashes. Additional near matches have 64-bit perceptual hashes within Hamming
distance 4 and grayscale RMSE at most 0.030 on $128\times128$ images scaled
to $[0,1]$. Counts are cross-manifest image pairs.}
\label{tab:app-image-overlap}
\begin{tabular}{@{}lrr@{}}
\toprule
Manifests compared & Exact matches & Additional near matches \\
\midrule
Stage1 and PD & 0 & 0 \\
Stage1 and DM & 0 & 0 \\
PD and DM & 144 & 0 \\
\bottomrule
\end{tabular}
\end{table}

\subsection{Patient folds and label definitions}
For the final PD system we used patient-level five-fold validation with split
seeds 42, 1042 and 2042. All images of a patient remained together in each
split. Validation labels were used for fold-local checkpoint selection and
evaluation; the fixed routing and voting rules use only candidate predictions.

\begin{table}[!htbp]
\centering\small
\caption{Patients eligible for each downstream comparison. Counts refer to distinct subjects before repeated validation.}
\label{tab:app-cohorts}
\begin{tabular}{@{}llrl@{}}
\toprule
Task & Input & Patients & Class counts \\
\midrule
Diagnosis & Paired IVCM & 110 & 47 controls, 63 PD \\
Diagnosis & Nerve measurements & 94 & 31 cataract controls, 63 PD \\
H\&Y & PD cohort & 63 & 22 / 30 / 11 in groups 1 / 2 / 3 \\
Subtype & PD cohort & 63 & 23 TD, 40 non-TD \\
DM transfer & IVCM & 80 & 20 controls, 60 DR \\
\bottomrule
\end{tabular}
\end{table}

H\&Y and subtype use separate patient partitions stratified by the corresponding task label.
All three H\&Y candidates share the same fold assignments; the five subtype
candidates likewise use a common partition. Each patient contributes one
prediction per repeat. In Stage1, single-region rows put the available image
into both input slots. Those rows can provide a central-view HGrad target,
whereas the regional contrast uses the 219 genuine pairs.

\FloatBarrier
\section{Baseline inputs and comparison protocols}
\label{app:external-fairness}

\subsection{Data and supervision available to each method}
ResNet18 used ImageNet initialization. SimCLR had access to both fold-local
PD images and CORN-1 self-supervision; Table~\ref{tab:pd-main-results}
reports the stronger predefined source for each task. UNet-Only classified
nerve maps produced by a CORN-1-trained segmentation model; UNet-Dual also
received the corresponding raw images. The GNN skeletonized the same maps.
MTL/MAE used CORN-1 images for auxiliary reconstruction. The episodic
prototype baseline formed its class supports from labeled training patients.

Nerve-LR used the 11 measurements listed in the main text. Measurements were
averaged per patient. Within each outer training fold, the imputer and scaler
were refitted in each inner split and the logistic-regression penalty was
selected by three-fold macro-F1. The DM baselines used ImageNet initialization
(ResNet18), a PD-side SimCLR backbone, or the shared MTL/MAE encoder. The
repository contains the task-specific baseline configurations.

\subsection{Image baseline training settings}
The following tables retain the task-specific training settings for the image
baselines. They also identify which models consume raw images, nerve maps, or
graphs. The outer splits follow the protocol of each baseline family.

\begin{table}[!htbp]
\centering
\caption{\textbf{ResNet18 direct CNN baseline parameters.} Final settings for the plain supervised dual-stream CNN baseline.}
\label{tab:app-baseline-ledger-resnet}
\footnotesize
\setlength{\tabcolsep}{5pt}
\begin{tabular}{@{}>{\raggedright\arraybackslash}p{0.27\linewidth}>{\raggedright\arraybackslash}p{0.66\linewidth}@{}}
\toprule
\textbf{Item} & \textbf{Final setting} \\
\midrule
Input / dataset & \texttt{pd\_dataset\_en}, \texttt{pd\_all}\newline central--whorl dual stream \\
Backbone / initialization & ResNet18; ImageNet \texttt{Half} initialization \\
Loss / head & CE with label smoothing; multi-dropout classification head \\
Augmentation & horizontal flip 0.5\newline rotation $\pm 6^{\circ}$\newline brightness/contrast 0.12 \\
Optimizer / scheduler & Adam or AdamW; cosine schedule \\
LR / WD & $2\times10^{-4}$ / $3\times10^{-3}$ \\
Epochs / patience & 40 epochs for \texttt{pd\_binary} and \texttt{pd\_stage}\newline 30 epochs for \texttt{pd\_subtype}\newline patience 6 \\
Evaluation / seeds & patient-level 5-fold $\times$ 3 repeats\newline seeds 42, 202, 3407 \\
Source experiments & P11--P13 for \texttt{pd\_binary}/\texttt{pd\_stage}\newline C02 for \texttt{pd\_subtype} \\
\bottomrule
\end{tabular}
\end{table}

\begin{table}[!htbp]
\centering
\caption{\textbf{SimCLR baseline parameters.} Contrastive pretraining and supervised fine-tuning settings for the reported SimCLR rows.}
\label{tab:app-baseline-ledger-simclr}
\footnotesize
\setlength{\tabcolsep}{5pt}
\begin{tabular}{@{}>{\raggedright\arraybackslash}p{0.18\linewidth}>{\raggedright\arraybackslash}p{0.38\linewidth}>{\raggedright\arraybackslash}p{0.38\linewidth}@{}}
\toprule
\textbf{Item} & \textbf{\texttt{pd\_binary} / \texttt{pd\_subtype}} & \textbf{\texttt{pd\_stage}} \\
\midrule
Architecture & dual-stream ResNet18 & dual-stream ResNet18 \\
Initialization & ImageNet \texttt{Half}\newline frozen backbone in fine-tuning & ImageNet \texttt{Half}\newline frozen backbone in fine-tuning \\
SSL source & Fold-local PD images & CORN-1 images \\
Pretrain epochs & 100 & 100 \\
Temperature & 0.4 & 0.5 \\
Projection dim / hidden & 256 / 256 & 128 / 256 \\
Pretrain LR / WD / batch & $10^{-3}$ / $10^{-4}$\newline batch 32 & $10^{-3}$ / $10^{-4}$\newline batch 32 \\
Fine-tune LR / WD & $2\times10^{-4}$\newline WD $5\times10^{-3}$ & $2\times10^{-4}$\newline WD $3\times10^{-3}$ \\
Backbone LR ratio & 0.0 & 0.0 \\
Fine-tune regularization & dropout 0.4\newline label smoothing 0.1 & dropout 0.4\newline label smoothing 0.1 \\
Batch / scheduler / patience & batch 16\newline warmup-cosine\newline patience 8 & batch 16\newline warmup-cosine\newline patience 8 \\
Final config & C304 / E900-C304 anchor & P41b/P42b/P43b \\
\bottomrule
\end{tabular}
\end{table}

\begin{table}[!htbp]
\centering
\caption{\textbf{UNet-derived CNN baseline parameters.} Settings for the UNet-map-only and raw-plus-map dual-view CNN baselines.}
\label{tab:app-baseline-ledger-unet}
\footnotesize
\setlength{\tabcolsep}{5pt}
\begin{tabular}{@{}>{\raggedright\arraybackslash}p{0.18\linewidth}>{\raggedright\arraybackslash}p{0.38\linewidth}>{\raggedright\arraybackslash}p{0.38\linewidth}@{}}
\toprule
\textbf{Item} & \textbf{UNet-Only} & \textbf{UNet-Dual} \\
\midrule
Input representation & \texttt{pd\_dataset\_unet\_en}\newline UNet-map central--whorl dual stream & raw image + UNet map\newline dual-view fusion \\
Backbone / initialization & ResNet18; \texttt{Half}\newline backbone LR ratio 0.03 & ResNet18; \texttt{Half}\newline shared backbone \\
Best config & C05\newline \texttt{resnet18\_ultra}\newline \texttt{reg\_conservative} & B04 for \texttt{pd\_binary}/\texttt{pd\_stage}\newline D01 for \texttt{pd\_subtype} \\
Fusion setting & single modality & fusion dim 256 for B04\newline fusion dim 384 for D01 \\
Loss & CE with label smoothing 0.12 & CE \\
LR / WD & $1.2\times10^{-4}$ / $6\times10^{-3}$ & $2\times10^{-4}$ / $5\times10^{-3}$ \\
Dropout & 0.45 & 0.3 classifier dropout \\
Epochs / batch & 45 / implementation default & 50 / 8 \\
Scheduler / patience & cosine\newline patience 8 & cosine\newline patience 10 \\
Evaluation & patient-level 5-fold $\times$ 3 repeats & patient-level 5-fold $\times$ 3 repeats \\
\bottomrule
\end{tabular}
\end{table}

\begin{table}[!htbp]
\centering
\caption{\textbf{GNN baseline parameters.} Graph-construction and message-passing settings for the skeleton-graph baseline.}
\label{tab:app-baseline-ledger-gnn}
\footnotesize
\setlength{\tabcolsep}{5pt}
\begin{tabular}{@{}>{\raggedright\arraybackslash}p{0.27\linewidth}>{\raggedright\arraybackslash}p{0.66\linewidth}@{}}
\toprule
\textbf{Item} & \textbf{Final setting} \\
\midrule
Input / graph source & \texttt{pd\_dataset\_unet\_en}\newline U-Net map $\rightarrow$ skeleton $\rightarrow$ graph \\
Node / edge features & 5 node features; 3 edge features \\
Architecture & central--whorl dual graph streams; shared encoder disabled \\
Main backbone & GAT, hidden 64, 2 layers, 4 heads \\
Subtype diagnostic variant & GraphSAGE C02 also recorded; main AUC uses C01 GAT \\
Loss & multi-task CE; missing labels masked \\
Optimizer / LR / WD & Adam; $10^{-3}$; $10^{-4}$ \\
Regularization & GNN dropout 0.2\newline classifier dropout 0.3\newline label smoothing 0.0 \\
Scheduler / warmup & cosine with 5 warmup epochs \\
Epochs / batch / patience & 80 / 16 / 15 \\
Evaluation / seeds & patient-level 5-fold $\times$ 3 repeats\newline P01/P02/P03 seeds 42/202/3407 \\
\bottomrule
\end{tabular}
\end{table}

\begin{table}[!htbp]
\centering
\caption{\textbf{MTL/MAE baseline parameters.} Settings for the main MAE-assisted multi-task model and subtype hierarchy variant.}
\label{tab:app-baseline-ledger-mtl}
\footnotesize
\setlength{\tabcolsep}{5pt}
\begin{tabular}{@{}>{\raggedright\arraybackslash}p{0.18\linewidth}>{\raggedright\arraybackslash}p{0.38\linewidth}>{\raggedright\arraybackslash}p{0.38\linewidth}@{}}
\toprule
\textbf{Item} & \textbf{G03 main MTL/MAE} & \textbf{I01 subtype hierarchy} \\
\midrule
Dataset / auxiliary source & \texttt{pd\_dataset\_en}\newline CORN-1 MAE images & \texttt{pd\_dataset\_en}\newline subtype-only hierarchy \\
Backbone / initialization & ResNet18; ImageNet \texttt{Half} & ResNet18; ImageNet \texttt{Half} \\
Branch A & MAE & MAE \\
Task weights & PD 0.40\newline stage 0.05\newline subtype 0.10\newline MAE 0.40 & subtype 1.0\newline PD/stage 0 \\
LR / WD & $10^{-4}$ / 0.10 & $2\times10^{-4}$ / $3\times10^{-3}$ \\
Backbone LR ratio & 0.05 & implementation default for I01 line \\
Dropout & 0.2 & 0.2 \\
Epochs / batch & 50 / 16 & 40 / 16 \\
Stabilization & mixed precision\newline gradient clipping 1.0 & data augmentation enabled \\
Scheduler / patience & cosine\newline patience 6 & fold-local early stopping \\
Evaluation & patient-level 5-fold $\times$ 3 repeats & patient-level 5-fold $\times$ 3 repeats \\
\bottomrule
\end{tabular}
\end{table}

\begin{table}[!htbp]
\centering
\caption{\textbf{Prototype baseline parameters.} Episodic training and fold-local prototype inference settings for the reported prototype rows.}
\label{tab:app-baseline-ledger-prototype}
\footnotesize
\setlength{\tabcolsep}{5pt}
\begin{tabular}{@{}>{\raggedright\arraybackslash}p{0.16\linewidth}>{\raggedright\arraybackslash}p{0.25\linewidth}>{\raggedright\arraybackslash}p{0.25\linewidth}>{\raggedright\arraybackslash}p{0.25\linewidth}@{}}
\toprule
\textbf{Item} & \textbf{\texttt{pd\_binary}} & \textbf{\texttt{pd\_stage}} & \textbf{\texttt{pd\_subtype}} \\
\midrule
Config ID & P11--P13 / E503 & P21--P23 / H816 & I900 \\
Task name & \texttt{pd} & \texttt{stage} & \texttt{pd\_subtype\_tdrest} \\
Architecture & dual-stream\newline ResNet18 Half & dual-stream\newline ResNet18 Half & dual-stream\newline ResNet18 Half \\
Embedding / query & 128 / 4 & 128 / 4 & 128 / 4 \\
Way / shot & 2 / 3 & 3 / 5 & 2 / 3 \\
Feature dropout & 0.70 & 0.55 & 0.65 \\
LR / WD & $5\times10^{-6}$\newline WD 0.03 & $5\times10^{-5}$\newline WD 0.01 & $10^{-5}$\newline WD 0.01 \\
Epochs / patience & 24\newline patience 2 & 35\newline patience 4 & 30\newline patience 3 \\
Episodes & 80 train episodes/epoch\newline 200 validation episodes & same & same \\
Evaluation & patient 5-fold $\times$ 3 seeds & patient 5-fold $\times$ 3 seeds & patient 5-fold $\times$ 3 repeats \\
\bottomrule
\end{tabular}
\end{table}

\subsection{Clinical measurement baseline}
The clinical measurements comprise seven central and four whorl-region nerve
metrics. The PD set contains four measurements for each of 63 patients.
The 31 cataract controls contribute 118 measurements. All measurements for a
patient are averaged before the outer split. The outer training fold selects
$C$ from $\{0.01,0.1,1,10\}$ and refits the selected pipeline on that fold.
The H\&Y and subtype evaluations follow the same patient partitions as
TopoDiSE for these tasks.

\FloatBarrier
\section{Training settings and candidate inference}
\label{app:topodise-budget}

\subsection{Three-stage optimization}
Stage0 trained on $224\times224$ images for 200 epochs (batch 16, learning
rate $10^{-4}$), with unit weight on image diffusion and 0.1 on mask BCE.
Stage1 ran for 80 epochs (batch 16, learning rate $2\times10^{-5}$), with
weights 0.5 on region contrast and 0.05 on dual-threshold HGrad BCE.
The frozen topology head passed gradients to the trainable encoder parameters.
Stage2 loaded the raw encoder weights and trained residual adapters, MIL
pooling and the task-specific heads. Fusion and projection were also trainable
for diagnosis and H\&Y, and fixed for subtype.

The semantic representation has 512 dimensions. The Stage0 pixel denoiser has a
1,000-step cosine noise schedule and conditioning dropout 0.1. Stage1 applies
the regional loss to central and whorl features with temperature 0.2. Its
HGrad target is computed from the central image. The selected checkpoint
supplies the raw encoder state to downstream candidates.

\subsection{Task-specific training and inference}
Each PD candidate trained for up to 50 epochs with patience 8. Diagnosis used
US0 with cross-entropy and morphology fusion. H\&Y candidates shared a
support-LSE rule; U1 and U2 used class-conditioned diffusion at augmentation
timesteps 45 and 80. Subtype candidates used class-weighted classification;
TDREST2, 5 and 7 enabled feature diffusion at timesteps 90, 70 and 90.
TDREST5 and 7 selected checkpoints by balanced patient accuracy, and the other
PD candidates by patient accuracy. Each candidate trained independently.
\begin{table}[!htb]
\centering\small
\caption{Fixed H\&Y and subtype candidates and inference rules. Values are fold means from the GT+HGrad main run.}
\label{tab:app-candidates}
\begin{tabular}{@{}llrr@{}}
\toprule
Candidate or rule & Configuration & Acc. & Macro-F1 \\
\midrule
U0 & No feature diffusion & 0.567 & 0.516 \\
U1 & Gentle, one support augmentation & 0.569 & 0.526 \\
U2 & Medium, two support augmentations & 0.551 & 0.519 \\
Mean probability & Average three probability vectors & 0.593 & 0.546 \\
Minimum entropy & Fixed H\&Y rule & 0.598 & 0.569 \\
\midrule
TDREST0 & Focal $\gamma=3$ & 0.678 & 0.508 \\
TDREST1 & Focal $\gamma=2$ & 0.720 & 0.608 \\
TDREST2 & CE, feature diffusion & 0.689 & 0.517 \\
TDREST5 & Focal $\gamma=1.5$, minority diffusion & 0.623 & 0.598 \\
TDREST7 & CE, diffusion, TD weight multiplier 1.5 & 0.548 & 0.516 \\
Mean probability & Average five probability vectors & 0.625 & 0.571 \\
Majority vote & Fixed subtype rule & 0.752 & 0.669 \\
\bottomrule
\end{tabular}
\end{table}

The residual adapters have rank 32 and the gated attention layer has hidden
dimension 128. Where morphology is enabled, 21 image statistics are projected
to 32 dimensions for late fusion. The H\&Y prototype embeddings have dimension
128, temperature 0.5 and loss weight 0.8. U0 uses real training-fold supports.
U1 and U2 augment these with one and two generated supports per augmentation
setting, respectively. Each candidate produces its own patient probability
vector. Minimum-entropy routing selects one of the three vectors for each
patient.

For subtype screening, TDREST0 and TDREST1 use focal exponents 3 and 2.
TDREST5 combines minority-class feature diffusion with a focal exponent of 1.5.
TDREST7 uses weighted cross-entropy and gives TD an additional weight factor
of 1.5. Each candidate contributes one vote; the probability average shown in
Table~\ref{tab:app-candidates} evaluates the same five outputs with a different
decision rule.

\FloatBarrier
\section{Patient-level results and encoder comparison}
\label{app:patient-level-results}
\subsection{Class-specific performance}
\begin{table}[!htb]
\centering\small
\caption{Pooled patient predictions over three repeats. Predicted columns follow the true-class order shown within each task. Each patient contributes once per repeat.}
\label{tab:app-confusion}
\begin{tabular}{@{}llrrrr@{}}
\toprule
Task & True class & Pred. 1 & Pred. 2 & Pred. 3 & Recall \\
\midrule
Diagnosis & Control & 115 & 26 & -- & 0.816 \\
 & PD & 23 & 166 & -- & 0.878 \\
H\&Y & Group 1 & 30 & 31 & 5 & 0.455 \\
 & Group 2 & 22 & 66 & 2 & 0.733 \\
 & Group 3 & 8 & 8 & 17 & 0.515 \\
Subtype & TD & 39 & 30 & -- & 0.565 \\
 & Non-TD & 17 & 103 & -- & 0.858 \\
\bottomrule
\end{tabular}
\end{table}

H\&Y group 1 had the lowest recall (0.455); group 2 had the highest (0.733).
For subtype, recall was 0.565 for TD and 0.858 for non-TD. These are
cohort-specific estimates under repeated patient-level validation.

\subsection{Fold-level results for the final PD system}
Table~\ref{tab:app-pd-folds} records the final patient predictions summarized
within each validation fold. The three blocks correspond to split seeds 42,
1042 and 2042. For H\&Y and subtype, the task-specific decision rule is applied
to each patient before computing the fold metric.
\begin{table}[!htbp]
\centering\small
\caption{GT+HGrad performance by fold for the three PD tasks. Each cell is accuracy / macro-F1.}
\label{tab:app-pd-folds}
\begin{tabular}{@{}rrccc@{}}
\toprule
Repeat & Fold & Diagnosis & H\&Y & TD/non-TD \\
\midrule
1 & 1 & 0.826 / 0.823 & 0.714 / 0.731 & 0.769 / 0.769 \\
1 & 2 & 0.826 / 0.826 & 0.462 / 0.431 & 1.000 / 1.000 \\
1 & 3 & 0.864 / 0.856 & 0.417 / 0.238 & 0.615 / 0.381 \\
1 & 4 & 0.952 / 0.951 & 0.750 / 0.735 & 0.750 / 0.697 \\
1 & 5 & 0.810 / 0.809 & 0.583 / 0.594 & 0.667 / 0.400 \\
\midrule
2 & 1 & 0.913 / 0.912 & 0.500 / 0.482 & 0.615 / 0.381 \\
2 & 2 & 0.913 / 0.913 & 0.846 / 0.869 & 0.769 / 0.764 \\
2 & 3 & 0.864 / 0.861 & 0.667 / 0.651 & 0.692 / 0.690 \\
2 & 4 & 0.762 / 0.741 & 0.500 / 0.481 & 0.917 / 0.899 \\
2 & 5 & 0.762 / 0.721 & 0.667 / 0.472 & 0.833 / 0.812 \\
\midrule
3 & 1 & 0.913 / 0.912 & 0.500 / 0.493 & 0.846 / 0.819 \\
3 & 2 & 0.913 / 0.909 & 0.615 / 0.648 & 0.769 / 0.764 \\
3 & 3 & 0.773 / 0.760 & 0.500 / 0.508 & 0.615 / 0.511 \\
3 & 4 & 0.857 / 0.857 & 0.833 / 0.780 & 0.667 / 0.400 \\
3 & 5 & 0.810 / 0.806 & 0.417 / 0.423 & 0.750 / 0.748 \\
\bottomrule
\end{tabular}
\end{table}

\subsection{Frozen encoders under a common Core Stage2}
In a separate experiment, frozen encoders were evaluated with the same patient
folds and a common Core Stage2 readout that omitted task-specific inference.
SimCLR and the topology-aware encoder used CORN-1 images; DINOv2-S/14 used
external LVD-142M pretraining. The comparison controls the downstream
protocol while pretraining resources differ.
\begin{table}[!ht]
\centering
\caption{\textbf{Frozen-encoder comparison under a common Core Stage2.}
Values are mean patient-level accuracy and macro-F1 over 15 folds. All encoders
use the same patient folds and common downstream protocol.}
\label{tab:app-core-stage2-encoders}
\scriptsize
\setlength{\tabcolsep}{3pt}
\begin{tabular}{@{}lcccccc@{}}
\toprule
\textbf{Encoder} & \multicolumn{2}{c}{\textbf{PD diagnosis}} & \multicolumn{2}{c}{\textbf{H\&Y}} & \multicolumn{2}{c}{\textbf{TD/non-TD}} \\
\cmidrule(lr){2-3}\cmidrule(lr){4-5}\cmidrule(lr){6-7}
& \textbf{Acc.} & \textbf{Macro-F1} & \textbf{Acc.} & \textbf{Macro-F1} & \textbf{Acc.} & \textbf{Macro-F1} \\
\midrule
SimCLR & 0.745 & 0.736 & 0.333 & 0.306 & 0.541 & 0.510 \\
DINOv2-S/14 & 0.679 & 0.657 & 0.447 & 0.368 & 0.528 & 0.484 \\
TopoDiSE & 0.758 & 0.750 & 0.450 & 0.353 & 0.577 & 0.528 \\
\bottomrule
\end{tabular}
\end{table}

\subsection{Image baseline fold records}
The next three tables give fold-level scores for the published image baselines.
Fold indices follow the archived order for each method. A common index across
different baseline families need not identify the same patient partition.

\begin{table}[!htbp]
\centering
\caption{\textbf{Baseline \texttt{pd\_binary} fold ledger.} Complete 15-fold patient-level accuracy and macro-F1 sequences for the image baselines.}
\label{tab:app-baseline-fold-binary}
\scriptsize
\setlength{\tabcolsep}{3pt}
\begin{tabular}{@{}lp{0.40\linewidth}p{0.40\linewidth}@{}}
\toprule
\textbf{Method} & \textbf{Accuracy by fold} & \textbf{Macro-F1 by fold} \\
\midrule
ResNet18 & 0.840, 0.708, 0.727, 0.850, 0.842, 0.880, 0.792, 0.727, 0.850, 0.684, 0.800, 0.833, 0.909, 0.900, 0.737 & 0.826, 0.695, 0.686, 0.847, 0.842, 0.879, 0.772, 0.686, 0.850, 0.661, 0.777, 0.812, 0.906, 0.896, 0.737 \\
SimCLR & 0.800, 0.833, 0.773, 0.850, 0.737, 0.840, 0.875, 0.682, 0.750, 0.789, 0.720, 0.667, 0.818, 0.750, 0.737 & 0.777, 0.829, 0.747, 0.840, 0.734, 0.833, 0.869, 0.576, 0.744, 0.789, 0.667, 0.664, 0.790, 0.715, 0.737 \\
UNet-Only & 0.760, 0.708, 0.773, 0.600, 0.632, 0.760, 0.583, 0.727, 0.750, 0.632, 0.680, 0.667, 0.682, 0.750, 0.737 & 0.724, 0.704, 0.760, 0.524, 0.632, 0.724, 0.556, 0.705, 0.749, 0.614, 0.632, 0.644, 0.681, 0.715, 0.734 \\
UNet-Dual & 0.692, 0.833, 0.696, 0.714, 0.800, 0.692, 0.750, 0.783, 0.762, 0.750, 0.654, 0.750, 0.739, 0.762, 0.900 & 0.685, 0.832, 0.673, 0.708, 0.792, 0.675, 0.750, 0.781, 0.760, 0.733, 0.653, 0.750, 0.739, 0.760, 0.899 \\
GNN & 0.789, 0.824, 0.765, 0.846, 0.769, 0.789, 0.824, 0.765, 0.846, 0.769, 0.789, 0.824, 0.765, 0.846, 0.769 & 0.441, 0.452, 0.433, 0.458, 0.435, 0.441, 0.452, 0.433, 0.458, 0.435, 0.441, 0.452, 0.433, 0.458, 0.435 \\
MTL/MAE & 0.864, 0.727, 0.636, 0.955, 0.923, 0.818, 0.682, 0.818, 0.864, 0.808, 0.909, 0.773, 0.864, 0.773, 0.885 & 0.861, 0.718, 0.581, 0.954, 0.921, 0.812, 0.618, 0.812, 0.863, 0.793, 0.909, 0.747, 0.861, 0.772, 0.876 \\
Prototype & 0.760, 0.792, 0.636, 0.700, 0.842, 0.760, 0.708, 0.727, 0.700, 0.684, 0.800, 0.833, 0.773, 0.750, 0.579 & 0.740, 0.782, 0.633, 0.697, 0.842, 0.756, 0.661, 0.705, 0.700, 0.676, 0.788, 0.829, 0.768, 0.733, 0.578 \\
\bottomrule
\end{tabular}
\end{table}

\begin{table}[!htbp]
\centering
\caption{\textbf{Baseline \texttt{pd\_stage} fold ledger.} Complete 15-fold patient-level accuracy and macro-F1 sequences for the image baselines.}
\label{tab:app-baseline-fold-stage}
\scriptsize
\setlength{\tabcolsep}{3pt}
\begin{tabular}{@{}lp{0.40\linewidth}p{0.40\linewidth}@{}}
\toprule
\textbf{Method} & \textbf{Accuracy by fold} & \textbf{Macro-F1 by fold} \\
\midrule
ResNet18 & 0.400, 0.429, 0.462, 0.545, 0.600, 0.400, 0.429, 0.154, 0.545, 0.400, 0.333, 0.357, 0.462, 0.545, 0.800 & 0.190, 0.200, 0.291, 0.235, 0.250, 0.190, 0.200, 0.111, 0.235, 0.190, 0.167, 0.175, 0.211, 0.235, 0.552 \\
SimCLR & 0.400, 0.571, 0.462, 0.545, 0.600, 0.533, 0.429, 0.462, 0.545, 0.500, 0.400, 0.500, 0.462, 0.545, 0.700 & 0.271, 0.413, 0.211, 0.235, 0.250, 0.389, 0.200, 0.211, 0.235, 0.222, 0.269, 0.372, 0.211, 0.235, 0.433 \\
UNet-Only & 0.400, 0.429, 0.462, 0.545, 0.700, 0.400, 0.500, 0.462, 0.545, 0.600, 0.400, 0.429, 0.462, 0.545, 0.600 & 0.190, 0.200, 0.211, 0.235, 0.492, 0.190, 0.322, 0.211, 0.235, 0.250, 0.190, 0.200, 0.291, 0.235, 0.250 \\
UNet-Dual & 0.400, 0.429, 0.385, 0.455, 0.300, 0.400, 0.571, 0.462, 0.364, 0.600, 0.400, 0.429, 0.385, 0.455, 0.600 & 0.190, 0.200, 0.250, 0.286, 0.154, 0.190, 0.489, 0.211, 0.237, 0.250, 0.190, 0.200, 0.185, 0.315, 0.250 \\
GNN & 0.400, 0.429, 0.462, 0.364, 0.600, 0.400, 0.429, 0.462, 0.545, 0.600, 0.400, 0.429, 0.462, 0.545, 0.600 & 0.190, 0.200, 0.211, 0.178, 0.413, 0.190, 0.200, 0.222, 0.235, 0.250, 0.190, 0.200, 0.211, 0.235, 0.250 \\
MTL/MAE & 0.500, 0.583, 0.333, 0.417, 0.200, 0.667, 0.500, 0.250, 0.083, 0.467, 0.500, 0.333, 0.250, 0.500, 0.400 & 0.319, 0.455, 0.167, 0.196, 0.111, 0.457, 0.222, 0.133, 0.056, 0.333, 0.235, 0.226, 0.244, 0.222, 0.190 \\
Prototype & 0.467, 0.357, 0.462, 0.545, 0.300, 0.467, 0.500, 0.308, 0.636, 0.400, 0.333, 0.429, 0.385, 0.273, 0.600 & 0.452, 0.376, 0.344, 0.333, 0.343, 0.333, 0.390, 0.222, 0.400, 0.190, 0.256, 0.343, 0.252, 0.229, 0.460 \\
\bottomrule
\end{tabular}
\end{table}

\begin{table}[!htbp]
\centering
\caption{\textbf{Baseline \texttt{pd\_subtype} TD/non-TD fold ledger.} Complete 15-fold patient-level accuracy and macro-F1 sequences for the image baselines.}
\label{tab:app-baseline-fold-subtype}
\scriptsize
\setlength{\tabcolsep}{3pt}
\begin{tabular}{@{}lp{0.40\linewidth}p{0.40\linewidth}@{}}
\toprule
\textbf{Method} & \textbf{Accuracy by fold} & \textbf{Macro-F1 by fold} \\
\midrule
ResNet18 & 0.615, 0.615, 0.615, 0.667, 0.667, 0.769, 0.538, 0.615, 0.667, 0.667, 0.769, 0.615, 0.615, 0.667, 0.667 & 0.606, 0.381, 0.381, 0.657, 0.400, 0.764, 0.350, 0.511, 0.400, 0.400, 0.707, 0.381, 0.381, 0.400, 0.400 \\
SimCLR & 0.692, 0.769, 0.769, 0.667, 0.667, 0.769, 0.615, 0.615, 0.667, 0.667, 0.692, 0.769, 0.769, 0.667, 0.667 & 0.675, 0.707, 0.707, 0.400, 0.400, 0.707, 0.381, 0.381, 0.400, 0.400, 0.675, 0.707, 0.764, 0.400, 0.400 \\
UNet-Only & 0.615, 0.692, 0.615, 0.667, 0.667, 0.615, 0.615, 0.615, 0.667, 0.667, 0.615, 0.615, 0.615, 0.667, 0.667 & 0.381, 0.567, 0.381, 0.400, 0.400, 0.381, 0.381, 0.381, 0.400, 0.400, 0.381, 0.381, 0.381, 0.400, 0.400 \\
UNet-Dual & 0.615, 0.615, 0.615, 0.667, 0.667, 0.615, 0.615, 0.615, 0.667, 0.667, 0.615, 0.769, 0.615, 0.667, 0.667 & 0.381, 0.381, 0.381, 0.400, 0.400, 0.381, 0.381, 0.381, 0.400, 0.400, 0.381, 0.707, 0.381, 0.400, 0.400 \\
GNN & 0.615, 0.615, 0.615, 0.667, 0.667, 0.615, 0.615, 0.615, 0.667, 0.667, 0.615, 0.615, 0.615, 0.667, 0.667 & 0.381, 0.381, 0.381, 0.400, 0.400, 0.381, 0.381, 0.381, 0.400, 0.400, 0.381, 0.381, 0.381, 0.400, 0.400 \\
MTL/MAE & 0.615, 0.769, 0.615, 0.667, 0.667, 0.615, 0.615, 0.538, 0.833, 0.667, 0.615, 0.692, 0.615, 0.750, 0.750 & 0.381, 0.707, 0.381, 0.400, 0.400, 0.381, 0.381, 0.350, 0.778, 0.400, 0.381, 0.567, 0.381, 0.733, 0.748 \\
Prototype & 0.385, 0.769, 0.538, 0.667, 0.583, 0.308, 0.615, 0.538, 0.500, 0.500, 0.692, 0.538, 0.462, 0.500, 0.500 & 0.278, 0.745, 0.458, 0.667, 0.556, 0.235, 0.575, 0.458, 0.333, 0.486, 0.675, 0.512, 0.448, 0.486, 0.500 \\
\bottomrule
\end{tabular}
\end{table}

\FloatBarrier
\section{Target combinations and component ablations}
\label{app:ablation-ledger}

\subsection{Stage0 and Stage1 supervision targets}
The target comparison uses the complete downstream system and the same
patient folds for all six combinations. With Stage1 HGrad fixed, Stage0 GT
gave the highest diagnostic and H\&Y scores; Stage0 HGrad gave the highest
subtype scores (Table~\ref{tab:app-target-matrix}).
\begin{table}[!htb]
\centering\small
\caption{Complete target matrix. Each cell reports mean accuracy / macro-F1 over 15 folds with pretraining seed 42. HGrad denotes the dual-threshold target.}
\label{tab:app-target-matrix}
\begin{tabular}{@{}llccc@{}}
\toprule
Stage0 & Stage1 & PD diagnosis & H\&Y & TD/non-TD \\
\midrule
GT & HGrad & 0.850 / 0.844 & 0.598 / 0.569 & 0.752 / 0.669 \\
GT & Frangi & 0.851 / 0.844 & 0.585 / 0.521 & 0.636 / 0.399 \\
HGrad & HGrad & 0.832 / 0.822 & 0.525 / 0.428 & 0.825 / 0.793 \\
HGrad & Frangi & 0.845 / 0.835 & 0.477 / 0.340 & 0.668 / 0.463 \\
Frangi & HGrad & 0.773 / 0.748 & 0.513 / 0.464 & 0.721 / 0.631 \\
Frangi & Frangi & 0.777 / 0.756 & 0.536 / 0.489 & 0.671 / 0.537 \\
\bottomrule
\end{tabular}
\end{table}

\subsection{Matched candidate-level component ablations}
An independent experiment varied one component or stage at a time, using its
own Full reference (US0, U1 and TDREST5). We computed each metric from all
out-of-fold patient predictions within a repeat, then averaged across three
repeats for each of three pretraining seeds (42, 43, 44). Table~\ref{tab:app-components}
therefore reports individual candidates under matched training and inference
conditions.
\begin{table}[!htb]
\centering\scriptsize\setlength{\tabcolsep}{3pt}
\caption{Candidate-level accuracy / macro-F1. A dash indicates that the intervention does not apply to the reference candidate.}
\label{tab:app-components}
\begin{tabular}{@{}llccc@{}}
\toprule
ID & Intervention & Diagnosis & H\&Y & Subtype \\
\midrule
M0 & Full & 0.851 / 0.846 & 0.545 / 0.524 & 0.637 / 0.625 \\
A01 & Skip Stage1 & 0.837 / 0.832 & 0.552 / 0.543 & 0.741 / 0.722 \\
A02 & Remove Stage0 image diffusion & 0.798 / 0.793 & 0.591 / 0.548 & 0.686 / 0.671 \\
A03 & Remove Stage1 HGrad loss & 0.851 / 0.842 & 0.591 / 0.562 & 0.658 / 0.650 \\
A04 & Remove Stage1 regional loss & 0.839 / 0.834 & 0.593 / 0.557 & 0.651 / 0.645 \\
A05 & Remove both topology losses & 0.766 / 0.742 & 0.547 / 0.500 & 0.667 / 0.644 \\
B05 & Shuffle Stage0 GT grid & 0.800 / 0.785 & 0.587 / 0.572 & 0.584 / 0.572 \\
B06 & Shuffle Stage1 HGrad grid & 0.853 / 0.844 & 0.556 / 0.487 & 0.624 / 0.600 \\
D01 & Remove residual adapters & 0.835 / 0.831 & 0.524 / 0.487 & 0.640 / 0.630 \\
D04 & Fine-tune full encoder & 0.743 / 0.721 & 0.596 / 0.531 & 0.570 / 0.512 \\
D05 & Mean MIL pooling & 0.848 / 0.843 & 0.543 / 0.520 & 0.656 / 0.630 \\
D07 & Remove morphology fusion & 0.842 / 0.837 & 0.541 / 0.518 & -- \\
D08 & Image morphology alone & 0.573 / 0.408 & 0.460 / 0.410 & 0.698 / 0.677 \\
D10 & Remove H\&Y prototype module & -- & 0.540 / 0.453 & -- \\
E01 & Remove Stage2 feature diffusion & -- & 0.526 / 0.476 & 0.679 / 0.660 \\
D15 & Remove class weights & -- & -- & 0.672 / 0.625 \\
D16 & Replace focal loss with weighted CE & -- & -- & 0.644 / 0.632 \\
\bottomrule
\end{tabular}
\end{table}

Without Stage0 image diffusion, diagnostic macro-F1 fell by 0.053 (95\%
paired interval $[-0.086,-0.022]$). Skipping Stage1 increased TDREST5 macro-F1
by 0.098 ($[0.042,0.153]$). These intervals resample patients within class,
with the same patient weights across models, seeds and repeats.

B05 and B06 permute grid cells within each image, using the same permutation
for both target channels. Shuffling the Stage0 mask grid lowered diagnostic
macro-F1 by 0.060 (paired 95\% interval $[-0.108,-0.018]$), while H\&Y
macro-F1 rose from 0.524 to 0.572.

\subsection{Contribution of the two corneal regions}
We compared central and whorl inputs on the same patients with both views.
The paired reference used both regions; each single-region variant supplied
only the named region to the encoder. All rows used the same candidate per task
and the same folds. The paired input had the highest diagnostic macro-F1.
Central-only input gave higher macro-F1 for U1 and TDREST5 in this comparison.

\begin{table}[!htbp]
\centering\small
\caption{Matched regional-input ablation. Each entry is accuracy / macro-F1.
All three rows use 110 diagnosis patients and 63 patients for each PD-only task.}
\label{tab:app-region-ablation}
\begin{tabular}{@{}lccc@{}}
\toprule
Regional input & Diagnosis (US0) & H\&Y (U1) & Subtype (TDREST5) \\
\midrule
Central and whorl & 0.852 / 0.847 & 0.531 / 0.505 & 0.637 / 0.627 \\
Central only (C03) & 0.820 / 0.815 & 0.563 / 0.547 & 0.704 / 0.681 \\
Whorl only (C04) & 0.823 / 0.817 & 0.480 / 0.405 & 0.640 / 0.566 \\
\bottomrule
\end{tabular}
\end{table}

\subsection{H\&Y support-bank replay}
To isolate the inference rule, we replayed each selected U1 checkpoint on
the same queries and support bank. Classifier-only inference (D11) gave
0.390 / 0.373 accuracy / macro-F1; replacing support-LSE with one
mean prototype per class (D12) gave 0.413 / 0.410. Removing only the
generated support entries (E06) gave 0.481 / 0.460,
compared with 0.545 / 0.524 for full U1.

D10 in Table~\ref{tab:app-components} trains U1 without its prototype module.
The replay rows use the model trained with that module and change only how
its predictions are computed. In particular, removing generated supports
leaves the real training-fold supports available to classify each query.

\subsection{Subtype imbalance}
The two-class breakdown in Table~\ref{tab:app-subtype-imbalance} explains why
higher TDREST5 accuracy need not mean better TD recognition. Removing class
weights increases non-TD recall while reducing TD recall. These rows evaluate
the TDREST5 candidate before the five-member vote.
\begin{table}[!htbp]
\centering\small
\caption{Mean out-of-fold class recalls for matched TDREST5 interventions. The aggregation follows Table~\ref{tab:app-components}.}
\label{tab:app-subtype-imbalance}
\begin{tabular}{@{}llrr@{}}
\toprule
ID & Intervention & TD recall & Non-TD recall \\
\midrule
M0 & Full TDREST5 & 0.729 & 0.583 \\
D15 & Remove class weights & 0.498 & 0.772 \\
D16 & Replace focal loss with weighted CE & 0.749 & 0.583 \\
E01 & Disable feature diffusion & 0.638 & 0.703 \\
\bottomrule
\end{tabular}
\end{table}

\FloatBarrier
\section{Frozen structural probes}
\label{app:topology-diagnostics}

\subsection{Mask and horizontal-gradient targets}
For a pooled nerve mask, each cell contains the fraction of annotated
pixels within its footprint. The presence channel marks cells with nonzero
density. The core channel marks densities above the image mean plus one sample
standard deviation. The two HGrad channels use the corresponding pooled
horizontal intensity changes, thresholded at their image mean and at their mean
plus one sample standard deviation. Figure~\ref{fig:app-targets} illustrates
both constructions on the same CORN-1 images. Bright tissue boundaries can
activate HGrad outside the annotated nerves.

\begin{figure}[!htbp]
\centering
\includegraphics[width=\linewidth]{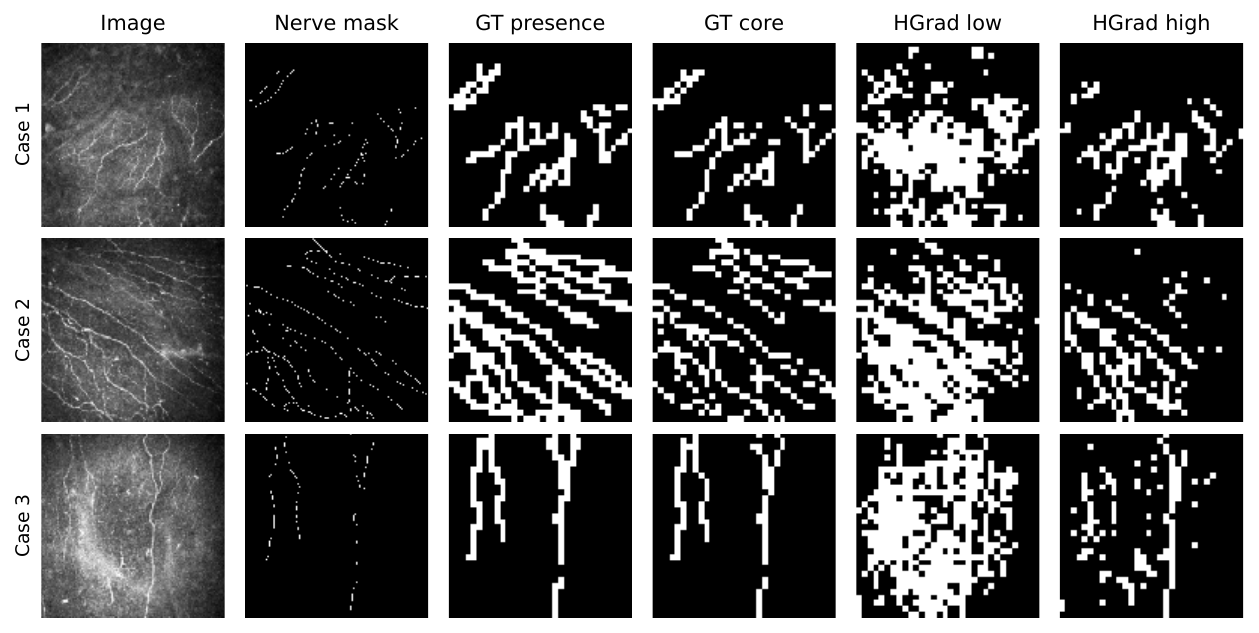}
\caption{Construction of the two supervision targets. Each row shows one
CORN-1 image, its annotation, both Stage0 GT channels and both Stage1 HGrad
channels. The three cases were chosen by identifier order before inspecting
probe predictions.}
\label{fig:app-targets}
\end{figure}

\subsection{Frozen grid readout and stage comparison}
We removed exact image duplicates from CORN-1 for probing and kept source
batches together, yielding 916 probe-training, 102 validation and 125 test
images. The encoder had been pretrained on CORN-1. For each frozen encoder,
we fitted a linear readout from the 512-dimensional embedding to the two
$32\times32$ mask-derived grids. Feature standardization used the probe-training
split and the validation split selected the probe epoch. We computed AP per
image and averaged over the same test images.
\begin{table}[!htb]
\centering\small
\caption{Stage0 probe AP for nerve presence / dense core. Each pair uses the same pretraining seed and probe split.}
\label{tab:app-absolute-probe}
\begin{tabular}{@{}lccc@{}}
\toprule
Seed & Without Stage0 GT & With Stage0 GT & Paired $\Delta$AP \\
\midrule
42 & 0.223 / 0.149 & 0.336 / 0.221 & +0.113 / +0.072 \\
43 & 0.221 / 0.147 & 0.328 / 0.213 & +0.108 / +0.066 \\
44 & 0.222 / 0.148 & 0.346 / 0.226 & +0.124 / +0.078 \\
\bottomrule
\end{tabular}
\end{table}

At seed 42, continuing from the GT-trained Stage0 encoder to the full
HGrad-regularized Stage1 encoder changed presence AP by $-0.0037$ and core AP
by $-0.0027$ on the same images. The two encoders had similar source-domain
grid readout at this seed.

The linear grid probe used AdamW (learning rate $10^{-3}$, weight decay
$10^{-4}$) for up to 200 epochs. Validation binary cross-entropy selected its
weights. Figure~\ref{fig:app-probe-predictions} shows continuous predictions
for the same cases as Figure~\ref{fig:app-targets}. For both encoders the
probability maps are spatially coarse, especially around fine branches.

\begin{figure}[!htbp]
\centering
\includegraphics[width=0.77\linewidth,height=0.65\textheight,keepaspectratio]{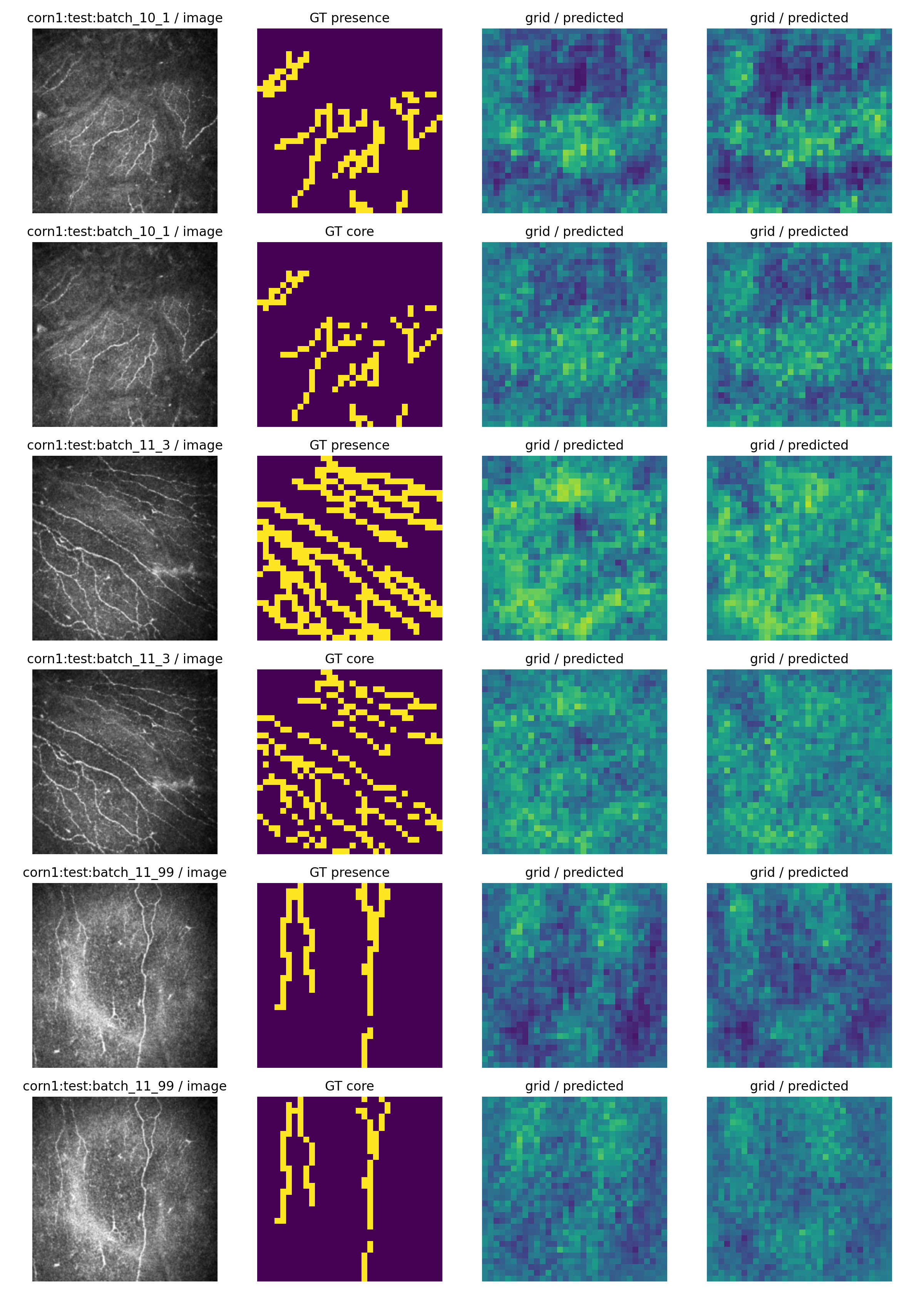}
\caption{Frozen linear-probe predictions at seed 42. From left to right:
input image, mask-derived grid, prediction from the GT-trained Stage0 encoder,
and prediction after Stage1. Each pair of rows shows presence and core for
the same image. Both prediction columns use the same probability scale.}
\label{fig:app-probe-predictions}
\end{figure}

\subsection{Recovery of skeleton descriptors}
A second probe regressed three scalar descriptors from each frozen feature
vector. Length density is the proportion of image pixels in the skeleton.
Branch and endpoint densities divide the respective pixel counts by skeleton
length. Feature and target scaling used the probe-training partition. We
selected the ridge penalty on validation images and measured mean absolute
error on the same 125 test images. The training-mean predictor provides a
reference for each descriptor in its own units.

\begin{table}[!htbp]
\centering\small
\caption{Skeleton-descriptor mean absolute error on the held-out probe split
(seed 42). Smaller values indicate closer agreement with the mask-derived descriptor.}
\label{tab:app-skeleton-descriptors}
\begin{tabular}{@{}lrrr@{}}
\toprule
Readout & Length density & Branch density & Endpoint density \\
\midrule
Training mean & 0.00477 & 0.00463 & 0.02932 \\
GT Stage0 & 0.00122 & 0.00405 & 0.02547 \\
After HGrad Stage1 & 0.00118 & 0.00428 & 0.02718 \\
\bottomrule
\end{tabular}
\end{table}

Both encoders recover length density more accurately than the training mean.
The errors for branch and endpoint density are closer to the reference error.
We also used each frozen encoder (seed 42) to retrieve the five training
images with the highest cosine similarity to each of the 125 test images.
We excluded images from the same source group or with the same image hash.
Mean intersection over union of the nerve-presence grids was 0.148 after
Stage0 and 0.139 after Stage1, compared with 0.122 for five randomly selected
eligible neighbors per query.

\FloatBarrier
\section{Reproducing the reported comparisons}
\label{app:reproducibility}
\subsection{Training entry points}
The repository linked in the abstract contains the three training stages
and the fixed Stage2 candidate recipes. Its English and Chinese guides are
organized by the tables in this paper.
\begin{table}[!htbp]
\centering\small
\caption{Public code and inputs for the reported PD experiments. Paths below
are relative to the repository root.}
\label{tab:app-repro}
\begin{tabular}{@{}p{.29\linewidth}p{.63\linewidth}@{}}
\toprule
Result & Entry point or artifact \\
\midrule
Final PD system & \path{experiments/topodise_experiments/batch_experiments.py}, option \texttt{--section main\_tasks} \\
Fixed candidate recipes & \path{experiments/topodise_experiments/recipes.py} \\
Full target matrix & \path{ablation_experiments/paper_targets.py} \\
Candidate interventions & \path{ablation_experiments/run.py}, option \texttt{--experiments paper B05 B06 D04.fixed D08} \\
Frozen probe split & \path{H01_H06_structure_diagnostics/build_manifest.py} in \path{ablation_experiments/} \\
Clinical baseline & \path{experiments/nerve_metrics_lr_experiments/paper.py} \\
\bottomrule
\end{tabular}
\end{table}

The main PD run uses one Stage0/Stage1 checkpoint chain at pretraining seed
42. It trains one diagnosis candidate, three H\&Y candidates and five subtype
candidates on each of 15 folds. The six-cell target matrix shares a Stage0
checkpoint across combinations that have the same Stage0 target. The
candidate-component experiments independently repeat pretraining at seeds 42,
43 and 44. For the Stage0 probes, the split manifest records the CORN-1 source
batch, image and mask for each example.

\subsection{Recovering the reported metrics}
For the main PD system, the evaluator aligns candidate predictions by patient
and fold, applies the task's fixed decision rule, then computes accuracy and
macro-F1 within each fold. Each fold receives weight $1/15$. The published
half-width for these PD scores is $1.96s/\sqrt{15}$, where $s$ is the sample
standard deviation of the fold scores. BAcc averages class recalls; the H\&Y
AUC averages the three one-versus-rest AUCs. The subtype AUC uses the
continuous mean probability and the confusion matrix uses the voted class.

For the component ablations, the five validation folds first form one
out-of-fold patient prediction set per repeat. We compute metrics on that
set and then average over repeats and pretraining seeds. The paired
intervals in Appendix~\ref{app:ablation-ledger} resample the same patients
for both models. The public scripts compute these metrics and reconstruct the
probe split from local CORN-1 files. Access to the clinical images and individual predictions
is restricted.

\FloatBarrier
\section{DM transfer protocol and class-wise performance}
\label{app:dm-transfer-details}

\subsection{Cohort and checkpoint}
The separate DMMIL3F1 transfer experiment used a Frangi-regularized Stage1
checkpoint and evaluated 20 controls and 60 DR subjects.
Each image was duplicated across the two encoder inputs because the cohort
lacks region labels. The Stage2 configuration used morphology fusion,
class-conditional feature diffusion ($T_{\mathrm{aug}}=60$), and balanced
patient accuracy for checkpoint selection. The archived AUC used continuous DR probabilities from the companion fold-logit run.

The 20 controls contribute 145 images and the 60 DR subjects contribute 469.
The positive label combines NDR and NPDR. Predicting the majority class for
every subject would give accuracy 0.750 and macro-F1 0.429 on this cohort.

\begin{table}[!ht]
\centering
\caption{Training configuration of the reported DMMIL3F1 transfer run.}
\label{tab:app-dm-topodise-config}
\small
\setlength{\tabcolsep}{4pt}
\begin{tabular}{@{}>{\raggedright\arraybackslash}p{0.29\linewidth}>{\raggedright\arraybackslash}p{0.65\linewidth}@{}}
\toprule
\textbf{Field} & \textbf{Value} \\
\midrule
Configuration & DMMIL3F1\_weak\_strict\_fdiff \\
Parent representation & US0 Stage1 parent with Frangi regularization \\
Task and split & dm\_binary, patient-level 5-fold $\times$ 3 repeats, 15 evaluated folds \\
Input construction & duplicated two-stream DM input, image size 224 \\
Mode and aggregation & Attention-MIL, patient aggregation by mean-logit patient output \\
Selection metric & balanced\_patient \\
Morphology side channel & enabled, 32-dimensional morphology embedding \\
Trainable transfer modules & adapters, fusion and projection, MIL pooling, morphology projection, classifier and feature-diffusion head \\
Optimizer schedule & learning rate $5\times10^{-4}$, weight decay 0.01, cosine schedule, warmup 3 epochs \\
Regularizers & classifier dropout 0.35, gradient clipping 1.0, label smoothing 0.05, class weights enabled \\
Feature diffusion & enabled; class conditional, soft-cosine gate, $T_{\mathrm{aug}}=60$, $k_{\mathrm{aug}}=1$ \\
Early stopping & patience 12, minimum delta 0.0001 \\
\bottomrule
\end{tabular}
\end{table}

\noindent\begin{minipage}{\linewidth}
\subsection{Class-wise transfer performance}
Across the three repeats, MTL/MAE recovered 30 of 60 control observations
and 165 of 180 DR observations. TopoDiSE recovered 55 and 140, respectively.
Figure~\ref{fig:app-dm-pooled-confusion} shows how the same overall accuracy
arose from different class-wise errors.
\par\medskip\centering
\includegraphics[width=\linewidth]{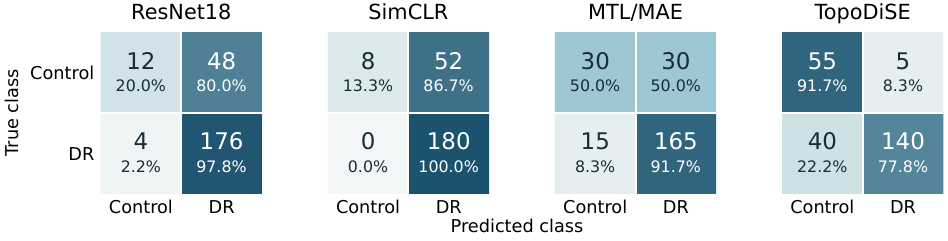}
\captionof{figure}{Patient-level DM transfer predictions pooled over 15 folds.
Cells give counts and percentages within each true class. Each of the 80
subjects appears once in each of three repeats.}
\label{fig:app-dm-pooled-confusion}
\end{minipage}

\begin{table}[ht]
\centering
\caption{\textbf{DM transfer baseline parameters.} Transfer and fine-tuning settings for non-TopoDiSE DM baselines.}
\label{tab:app-baseline-ledger-dm}
\footnotesize
\setlength{\tabcolsep}{5pt}
\begin{tabular}{@{}>{\raggedright\arraybackslash}p{0.16\linewidth}>{\raggedright\arraybackslash}p{0.25\linewidth}>{\raggedright\arraybackslash}p{0.25\linewidth}>{\raggedright\arraybackslash}p{0.25\linewidth}@{}}
\toprule
\textbf{Item} & \textbf{ResNet DM} & \textbf{SimCLR $\rightarrow$ DM} & \textbf{MTL $\rightarrow$ DM} \\
\midrule
Dataset & \texttt{dm\_dataset\_en} & \texttt{dm\_dataset\_en} & \texttt{dm\_dataset\_en} \\
Transfer source & ImageNet Half only & PD SimCLR backbone & PD MTL-MAE encoder \\
Architecture & dual-stream\newline ResNet18 & dual-stream\newline ResNet18 & ResNet18 encoder transfer \\
LR / WD / bb ratio & $2\times10^{-4}$ / $3\times10^{-3}$\newline ratio 0.05 & $2\times10^{-4}$ / $3\times10^{-3}$\newline ratio 0.1 & $3\times10^{-4}$ / $3\times10^{-3}$\newline ratio 0.05 \\
Epochs / batch & 40 / 8 & 40 / 16 & 30 / default \\
Regularizers & augmentation enabled\newline patience 6 & dropout 0.4\newline label smoothing 0.1\newline patience 8 & dropout 0.2 \\
Scheduler & cosine & warmup-cosine & fold-local fine-tuning schedule \\
Evaluation / seeds & patient 5-fold\newline seeds 42/202/3407 & patient 5-fold\newline seeds 42/202/3407 & patient 5-fold\newline seeds 42/202/3407 \\
Source config & \texttt{P2\_resnet18}\newline \texttt{dm\_patient\_acc} & \texttt{P3\_dmt0}\newline \texttt{dm\_patient\_acc} & \texttt{P2\_dmt2}\newline \texttt{dm\_patient\_acc} \\
\bottomrule
\end{tabular}
\end{table}

\FloatBarrier
\section{Clinical labels and interpretation}
\label{app:clinical-boundary}
\subsection{Endpoints and control composition}
The PD cohort follows Li et al.~\cite{li2025cornea}. H\&Y grouping uses three
cohort-provided severity groups. The subtype endpoint distinguishes 23 TD
patients from 30 PIGD and 10 mixed cases pooled as non-TD. These labels define
which cases contribute to each task and how to read its class-specific errors.

The recruited control cohort includes 31 cataract controls and 20 healthy
controls. Four healthy controls lack central views and fall outside the
paired-image diagnosis analysis. Clinical nerve measurements are available
for the cataract controls. Consequently, comparing the diagnosis accuracies
of Nerve-LR and the image system also changes the control population.

\noindent\begin{minipage}{\linewidth}
\subsection{Class-specific errors and clinical scope}
Figure~\ref{fig:app-clinical-boundary} summarizes the patient-level recalls
alongside the recruited control composition. Among H\&Y group~1 observations,
31 of 66 were assigned to group~2 across repeats. For subtype screening,
30 of 69 TD observations were classified as non-TD. The errors are relevant
when interpreting the corresponding overall accuracies.

\par\medskip
\centering
\includegraphics[width=\linewidth]{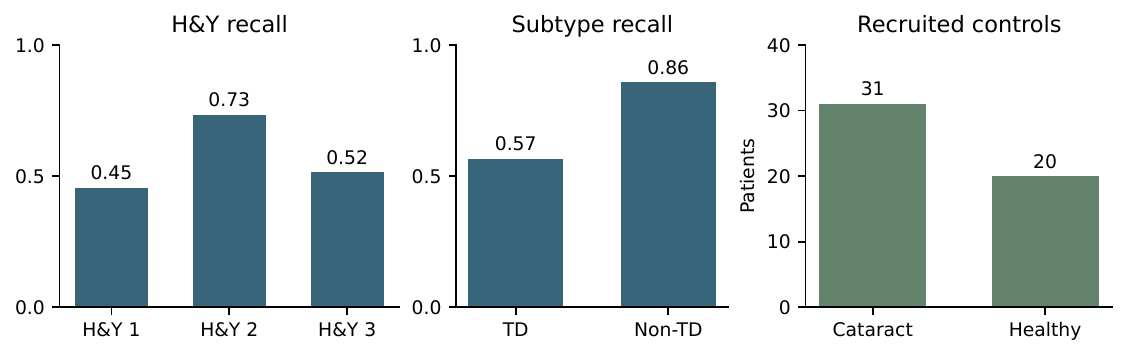}
\captionof{figure}{Cohort composition and class-specific behavior. H\&Y and subtype
recalls use the final GT+HGrad predictions pooled over three repeats.
The control panel counts distinct recruited subjects before image pairing.}
\label{fig:app-clinical-boundary}
\end{minipage}

The source study describes control matching and clinical exclusion
criteria~\cite{li2025cornea}. The files available for modelling contain task
labels and region assignments. Acquisition session, disease duration,
medication and ocular comorbidity were not included as adjustment variables in
the patient models. Independent validation with these covariates would help
distinguish disease-related image features from differences in acquisition or
control composition.

\clearpage
\fi
\bibliographystyle{splncs04}
\bibliography{main}
\end{document}